\documentclass{emulateapj}
\usepackage{graphicx}
\usepackage[usenames,dvipsnames]{color}

\shorttitle{Polarimetric Survey of Sh 2-29}
\shortauthors{Santos et al.}

\begin{document}

\title{
Optical/Near-IR Polarization Survey of Sh 2-29: Magnetic Fields, \\ 
Dense Cloud Fragmentations and Anomalous Dust Grain Sizes
\footnote{Based on observations collected at the, 
 National Optical Astronomy Observatory (CTIO, Chile) and Observat\'orio
 do Pico dos Dias, operated by Laborat\'orio Nacional de Astrof\'\i sica 
 (LNA/MCT, Brazil).}}

%
%
%
%

\author{F\'abio P. Santos\altaffilmark{1}, 
Gabriel A. P. Franco\altaffilmark{1}, 
Alexandre Roman-Lopes\altaffilmark{2}, \\ 
Wilson Reis\altaffilmark{1} \&
Carlos G. Rom\'an-Z\'u\~niga\altaffilmark{3}
}

\affil{\altaffilmark{1}Departamento de F\'isica -- ICEx -- UFMG, Caixa Postal 702, 30.123-970 \\
Belo Horizote, MG, Brazil; [fabiops;franco;wilsonr]@fisica.ufmg.br}
\affil{\altaffilmark{2}Departamento de Fisica -- Universidad de La Serena, Cisternas 1200, La Serena, Chile; roman@dfuls.cl}
\affil{\altaffilmark{3}Instituto de Astronom\'ia -- Universidad Nacional Aut\'onoma de M\'exico, Unidad Académica en Ensenada, \\
Ensenada BC 22860, Mexico; croman@astrosen.unam.mx}

{\footnotesize Accepted for The Astrophysical Journal}

\begin{abstract} 
Sh 2-29 is a conspicuous star-forming region marked by the presence of massive embedded stars 
as well as several notable interstellar structures.
In this research, our goals were to determine the role of magnetic fields and to study the size distribution
of interstellar dust particles within this turbulent environment.
We have used a set of optical and near-infrared polarimetric data obtained at OPD/LNA (Brazil)
and CTIO (Chile), correlated with extinction maps, 2MASS data and images from DSS and {\it Spitzer}.
The region's most striking feature is a swept out interstellar cavity whose polarimetric maps
indicate that magnetic field lines were dragged outwards, pilling up along its borders. 
This led to a higher magnetic strength value ($\approx400\,\mu$G) and an abrupt increase in polarization degree, 
probably due to an enhancement in alignment efficiency. Furthermore, dense cloud fragmentations 
with peak $A_{V}$ between $20$ and $37$ mag were probably triggered by its expansion. 
The presence of $24\,\mu$m point-like sources indicates possible newborn stars 
inside this dense environment. A statistical analysis of the angular dispersion function 
revealed areas where field lines are aligned in a well-ordered pattern, 
seemingly due to compression effects from the H{\sc ii} region expansion. 
Finally, Serkowski function fits were used to study the ratio of the total-to-selective extinction, 
reveling a dual population of anomalous grain particles sizes. This trend 
suggests that both effects of coagulation and fragmentation of interstellar grains are present in the region.

\end{abstract}


\section{Introduction}
\label{introduction}

It is well known that a large-scale magnetic field structure pervades the disk of the 
Milky Way Galaxy, including molecular clouds and star-forming regions 
\citep{mathewson1970,novak1989,novak1997,heiles2000,fosalba2002,heiles2005,santos2011}. 
It is possible that the presence of these magnetic field lines play an important role
on the initial stages of star-forming processes, providing support against the 
molecular cloud contraction. 
However, there is no general consent on whether these fields are important for the 
formation of sub-structures such as interstellar filaments, dense cores, and stars.
On one hand, if magnetic fields are dynamically effective on this kind of environment, then 
star formation is probably regulated by the process known as ambipolar diffusion 
\citep{mestel1956,nakano1979,mou1981,shu1987,lizano1989,heitsch2004,girart2009}, 
or by the diffusion processes related to magnetic reconnection \citep{leao2012}.
On the other hand, if magnetic fields do not exert a very important role on these
regions, then other effects, such as turbulence and stellar winds, must provide
support against the collapse of molecular clouds \citep{padoan2004,crutcher2005,mckee2007}.

Furthermore, on later stages, there is clearly a very dynamical interaction between the 
newborn embedded stellar populations and their surrounding interstellar medium.
This interaction significantly affects the morphology of gas and dust structures through radiation pressure 
and strong stellar winds, especially toward sites of massive star formation.
Since the magnetic field is generally frozen within the interstellar gas, its lines
are also critically affected in this process. The distortion of magnetic field lines
due to such interaction, have been detected near some Galactic star-forming environments
\citep[e.g., ][]{novak2000,matthews2002,kandori2007,kusakabe2008,tang2009,
santos2012}. 
However, several open questions still remain, related to the importance of magnetic 
fields in determining the sizes of H{\sc ii} regions, on the formation of radiation-driven filaments, 
and on the process of triggered star formation \citep{krumholz2007,arthur2011}.

The investigation of these questions is crucial to elucidate the underlying mechanisms 
controlling star formation. Unfortunately, when considering the widely different physical conditions
of star-forming regions throughout the Galaxy, only few observations of magnetic 
fields are available. Therefore, in order to evaluate its general importance, 
it is pivotal to study this feature from stellar to Galactic scales. Such analyses should 
serve as essential tests to provide additional constraints to several star formation models.

The best tool to map the sky-projected lines of magnetic fields is obtained from interstellar polarization 
by aligned dust grains. Typically, due to the dissipation of internal energy from the interaction
with the magnetic fields, the grains will be oriented with its major axis perpendicular to the 
field lines \citep{davis1951}. Therefore, if stellar light goes through 
such dichroic interstellar environment, the beam will be linearly polarized following
the magnetic field lines. This absorption effect is best observed using 
optical and near-infrared (near-IR) spectral bands. 

Observations of several distinct spectral bands allow a particularly detailed analysis 
of magnetic fields, as well as the prevailing grain alignment mechanisms. 
Furthermore, by studying polarization as a function of wavelength it is possible 
to compute the distribution of grain sizes, which directly affects the interstellar extinction law
\citep{gehrels1965,coyne1974,serkowski1975,codina1976,wilking1980,messinger1997}.

In this study we have focused on a rich star-forming region of the Galaxy, where 
the general aspects suggest a strong interaction between the stellar
population and the magnetic field lines. In Section \ref{s:descriptionsh229}
we give a detailed description of this area, along with a listing 
of the majority of observations conducted toward its direction, available in the literature.
Subsequently, Section \ref{s:obsdatash229} describes the observational methods
and the polarimetric data. Sections \ref{s:resultssh229} and 
\ref{s:analysisdiscussionsh229} presents the results, analysis 
and discussion. Finally, the conclusions are shown in Section \ref{s:conclusionssh229}.

\section{The Sh 2-29 star-forming region}
\label{s:descriptionsh229}

   \begin{figure*}[!t]
   \centering
   \includegraphics[width=\textwidth]{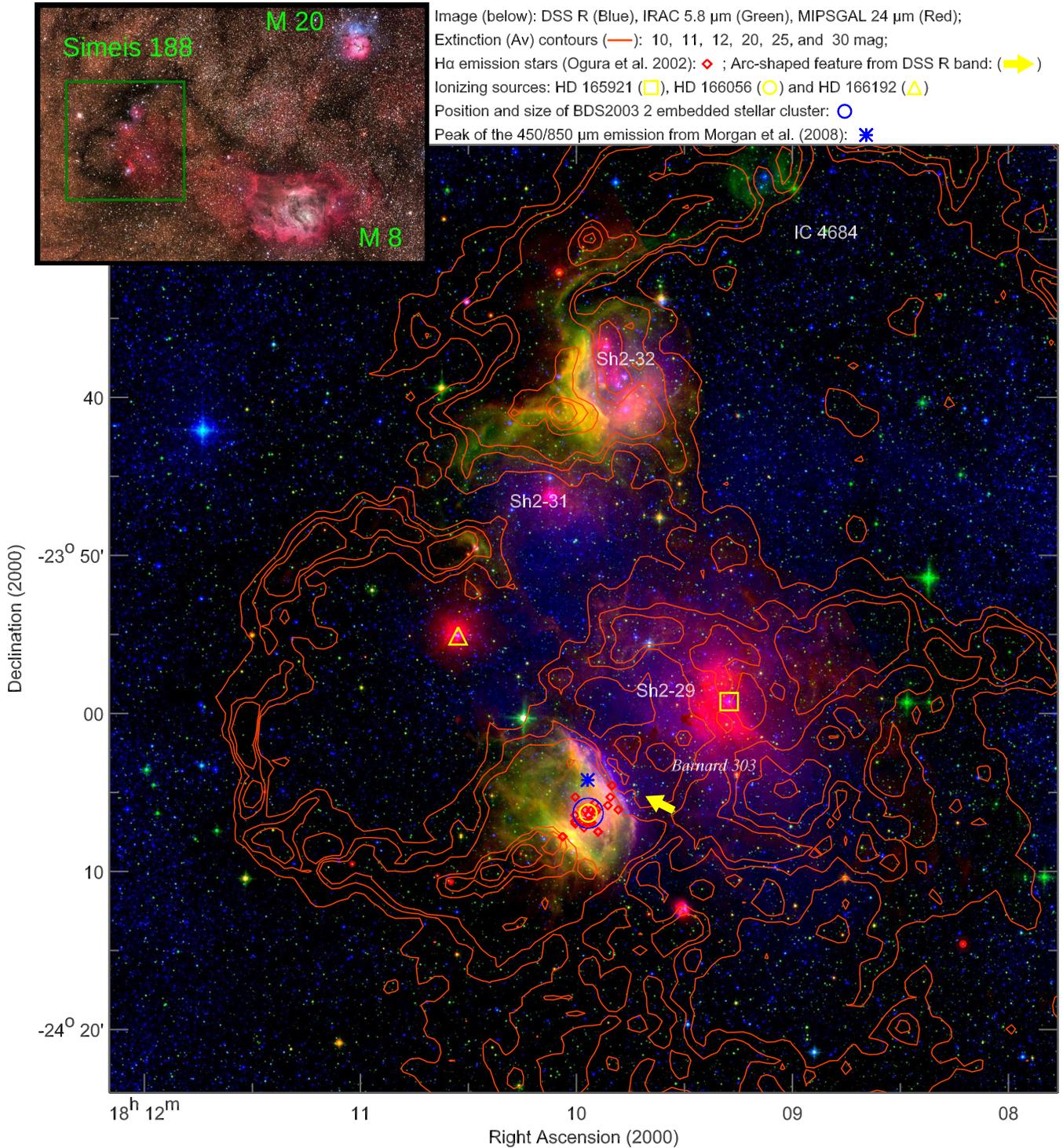} \\
      \caption{Image of the Simeis 188 region (which includes Sh 2-29), represented by a combination of 
               the following spectral bands: DSS R (blue), IRAC $5.8\,\mu$m (green), and MIPS $24\,\mu$m (red). 
               The smaller image at the top shows a wider optical view of the region, also encompassing 
               the M20 and M8 nebulae (source: {\tt http://www.nasaimages.org}). 
               Several stellar and interstellar features are marked, as specified by the indications 
               at the top, and also detailed in Section \ref{s:descriptionsh229}. The extinction contours
               shown by the orange lines was obtained by using the 2MASS data, as described
               in Section \ref{s:obsdatash229}.
              }
         \label{ism_sh229}
   \end{figure*}

The Sh 2-29 region was initially identified by \citet{sharpless1959}, and 
classified as an irregular shape H{\sc ii} region, displaying amorphous and 
filamentary structures (it is also known as NGC 6559 or IC 4685). 
This region is placed at the Simeis 188 Galactic complex, 
which is a rich star forming region at Sagittarius (near the direction of the 
Galactic center: $l\sim7\degr, b\sim-2\degr$), including several bright nebulae, 
such as IC 1275 (Sh 2-31), IC 1274 (Sh 2-32) and IC 4684 (see Figure \ref{ism_sh229}). 
According to \citet{herbig1957}, Sh 2-29 is probably associated to the same 
complex of M8 (the Lagoon Nebula) and M20 (the Trifid Nebula).

Figure \ref{ism_sh229} is a combination of DSS R band (blue), {\it Spitzer} IRAC $5.8 \,\mu$m 
(green), and {\it Spitzer} MIPS $24 \,\mu$m (red) 
\footnote{The data from {\it Spitzer} have been retrieved from the NASA/IPAC Infrared Science
Archive (IRSA). IRAC data are from the GLIMPSE survey ({\tt http://www.astro.wisc.edu/sirtf/}), 
while MIPS data are from the MIPSGAL survey ({\tt http://mipsgal.ipac.caltech.edu/index.html}).}.
Also overlaid on the image are the contours indicating the visual 
extinction $A_{V}$ (orange color, see Section \ref{s:obsdatash229}).
This area exhibits several interesting interstellar features, such as reflexion
nebulae, dark clouds (e.g., the serpent-shaped structure Barnard 303), and H$\alpha$
emission from the ionized gas (which may be noticed, for example, as extended structures 
in blue -- the R-band DSS image). The general interstellar morphology suggests a strong 
interaction between the embedded stellar population and the gas and dust structures.

\citet{sharpless1959} proposed that the H{\sc ii} region's main ionizing sources 
are the binary system HD165921 (spectral types O7{\sc v}+O9{\sc v}), HD166192 (B2{\sc ii}),
and HD166056 (B4{\sc i}/{\sc ii}), which are all relatively luminous stars in the visual band
(V = 7.4, 8.5, and 9.7 mag, respectively). The $24 \,\mu$m observations (red, Figure \ref{ism_sh229})
show a strong extended emission mainly around HD165921 and HD166192, 
indicating a heating of the surrounding interstellar material (particularly near 
HD165921, an arc-shaped structure may be noticed).

Thermal emission from the radio continuum (2695 GHz), 
which is typically related to free-free emission, was detected toward Sh 2-29 by
\citet{altenhoff1970}. Other radio observations were carried out aiming at obtaining 
radial velocities based on CO lines \citep{blitz1982}. Assuming a Galaxy rotation model \citep{fich1984},
a distance of $d=1.3\pm0.5$ kpc was computed.
CO-line studies have also been carried out by \citet{yamaguchi1999}, who interpreted the entire Sh 2-29 interstellar 
complex as an expanding shell. By assuming an expansion
velocity of 5 km\,s$^{-1}$ and a 5-10 pc radius, a dynamical age of 1-2 million years was estimated. 
Other studies of Sh 2-29 include far ultraviolet observations \citep{carruthers1984}, as well as
analysis of H$\alpha$ spectra \citep{fich1990}.

Although Sh 2-29 seems related to other nearby H{\sc ii} regions, the presence of different CO velocities 
suggests different distances. In fact, for M8 and M20, recent distance estimates suggest respectively
1250 and 1670 pc \citep{arias2007,rho2008}, although variations in the extinction law toward the Trifid
Nebula (M20) may indicate a somewhat higher distance of $2.7 \pm 0.5$ kpc \citep{cambresy2011}. Toward 
Sh 2-32, which is positioned $\approx$25 arcmin North of Sh 2-29 (and is supposedly part of the same 
interstellar complex), \citet{dahm2012} have determined a distance of $1.8 \pm 0.3$ kpc for the associated 
bright stellar cluster, which is consistent with other photometric and kinematic distance indicators
\citep{georgelin1973,vogt1975,fich1984,oka1999}.

By using data from the 2MASS near-IR survey \citet{bica2003} found an embedded stellar cluster
surrounding the ionizing source HD166056, denominated BDS2003 2 (with an angular dimension of $1.9'$). 
The cluster is at the center of a prominent cavity only seen in the mid/far-IR images 
(green/yellow emission from Figure \ref{ism_sh229}). Around this cavity, a large arc-shaped 
interstellar shell is visible in the DSS R-band image, and therefore is likely an expanding 
ionization front (assuming such feature is mainly due to the H$\alpha$ emission).
\citet{ogura2002} observed 23 H$\alpha$-emission sources toward this cluster, 
a feature which is typical of pre-main sequence objects (e.g., T Tauri or Herbig Ae/Be stars).

\citet{morgan2008} conducted 
SCUBA\footnote{{\it Submillimetre Common User Bolometer Array}, mounted on the
{\it James Clerk Maxwell Telescope} (Mauna Kea, Hawaii/EUA).} 
sub-mm observations toward Sh 2-29 (450 and 850 $\,\mu$m), 
revealing an extended dust emission feature $\approx$2 arcmin North of the BDS2003 2 cluster's center, 
coinciding with the position of the IRAS 18068-2405 source. \citet{urquhart2009} carried out
CO observations complemented with MSX and Spitzer data, concluding that no mid-IR point source may
be detected toward this feature, and therefore the region is still in a very early evolutionary stage, 
with most of its luminosity still in the sub-mm range. Furthermore, it is suggested
that star formation activity at this interstellar structure has possibly been externally triggered. 
\citet{urquhart2009} concluded that its current evolutionary status corresponds to the 
phase immediately after the emergence and exposition to an external ionization front, and before the arising 
of the first proto-stellar objects with significant mid-IR emission.

The only known polarimetric study of a region close to Sh 2-29 was conducted by \citet{mccall1990} 
toward M8, which is located at approximately $1.6^{\circ}$ from it. 
They have found a significantly anomalous value of the total-to-selective extinction ratio
of $R=4.6\pm0.3$, which is probably due to effects of evaporation and coagulation of interstellar grain
particles. Moreover, according to the interpretation of  \citet{mccall1990}, the overall cloud's morphology 
suggests a previous gravitational collapse following the magnetic field lines.

\section{Observational data}
\label{s:obsdatash229}

The linear polarization data used in this work are a combination of 
observations conducted at the Cerro Tololo Interamerican Observatory 
(CTIO, Chile), using the 0.9\,m telescope, and also at the 
Observat\'orio Pico dos Dias/Laborat\'orio Nacional de Astrof\'isica 
(OPD/LNA, Brazil), using both the 0.6\,m and the 1.6\,m telescopes.
The optical data (V, R, and I Johnson-Cousins bands) were obtained 
with the CTIO-0.9\,m and OPD-0.6\,m telescopes, while the near-IR data (H band) were
collected with the OPD-1.6\,m telescope. 

A brief description of the instrument and the data reduction process is given here. For 
a complete and detailed discussion, see \citet{santos2012}.
The polarimetric modules used in this work, from both OPD and CTIO, 
basically consist of a sequence of optical elements positioned in the 
stellar light path, before the beam reaches the detector. 
Assuming that the polarization orientation is defined by an angle $\theta$ relative to the 
North Celestial Pole, initially a half-wave plate causes a rotation of
the polarization plane. The plate itself may be rotated in discrete steps
of $22.5\degr$, resulting in consecutive polarization plane rotations 
of $45\degr$. The beam is subsequently split in two orthogonally polarized components, 
after passing through a Savart plate analyser. Lastly, each beam passes through a 
spectral filter and reaches an image detector. Therefore, the simultaneous 
analysis of the components' relative intensity, at all positions $\psi_i$ of the 
half-wave plate, may be used to adjust an oscillating cos$4\psi_i$ modulation function, 
which provides the $Q$ and $U$ Stokes parameters. Finally, polarization 
degree ($P$) and angle ($\theta$) are obtained through the standard relations:

\begin{equation}
P=\sqrt{Q^{2}+U^{2}} \ \ \ \mathrm{and} \ \ \ \theta=\frac{1}{2}\arctan({U/Q})
\label{e:polparameters}
\end{equation}

Image reduction and photometry were performed using IRAF\footnote{IRAF is distributed by the 
National Optical Astronomy Observatories, which are operated by the Association of 
Universities for Research in Astronomy, Inc., under cooperative agreement with the National 
Science Foundation.} routines \citep{tody1986}, especially from the DAOPHOT package.
Typical procedures were aimed at correction of bias level and flat-field pattern, 
sky subtraction (only for near-IR images), detection of point-like sources with counts
$5\sigma$ above the local background, and flux measurements using aperture photometry 
(PHOT task). 

Polarimetric parameters for all sources detected within each mapped field 
were computed from the ratio in measured flux for each beam using a set 
of specially designed IRAF routines \citep[PCCDPACK package,][]{pereyra2000}.
Calibration of polarization zero-point angle and degree was based on observations 
of polarimetric standards from several catalogs 
\citep{wilking1980,wilking1982,tapia1988,turnshek1990,larson1996}. Exposure times consisted
of $300$ and $600$ seconds, respectively for optical and near-IR observations, at 
each position of the half-wave plate. Stellar objects affected by cosmic rays, bad pixels, 
saturation or beam superposition due to crowding were removed, and only objects with 
$P/\sigma_{P} > 2$ were used in the subsequent analysis.
Furthermore, since $P$ is a positive quantity, it is usually overestimated 
through a simple application of Equation \ref{e:polparameters}. Therefore, in order to 
avoid this bias, the correction $P \rightarrow \sqrt{P^{2}-\sigma_{P}^{2}}$ was adopted, 
as suggested by \citet{wardle1974}.

   \begin{figure}[!t]
   \centering
   \includegraphics[width=0.48\textwidth]{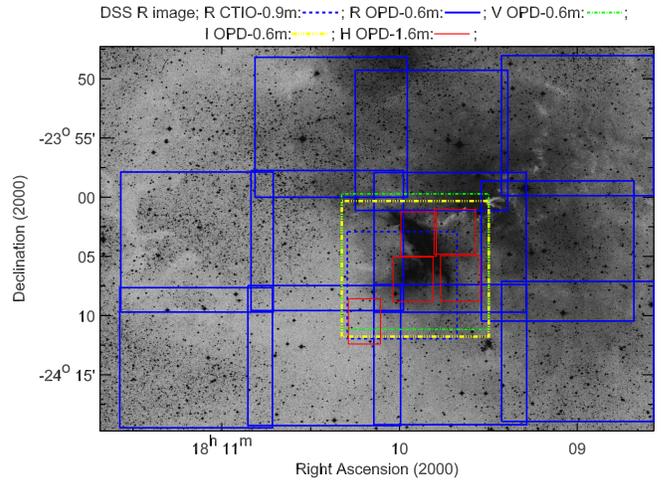} \\
      \caption{Position of the polarimetric mapped fields superposed to a R-band DSS image of Sh 2-29.
               Each line color and style is related to different spectral bands and telescopes
               used, as indicated by the caption at the top.
              }
         \label{obsfields_sh229}
   \end{figure}
%

   \begin{figure}[!t]
   \centering
   \includegraphics[width=0.48\textwidth]{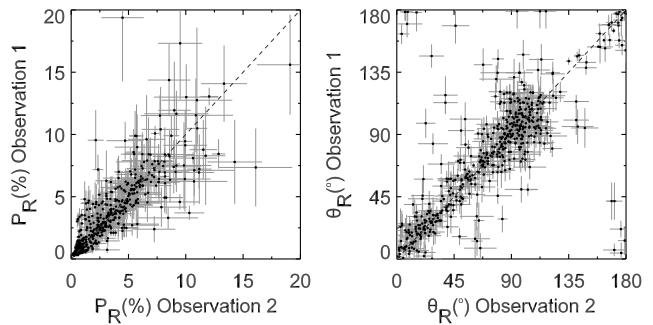} \\
      \caption{Comparison between polarization degrees ({\it left}) and polarization angles ({\it right})
               for Sh 2-29 objects observed more than once at R-band observations, due 
               to superposition of adjacent frames (see Figure \ref{obsfields_sh229}). 
               Equality lines (dashed) are included in order to facilitate the comparison.
              }
         \label{comparison_pol_sh229}
   \end{figure}

\begin{table*}
\centering
\caption{\label{tab_pol_sh229} Polarimetric data set for Sh 2-29 at optical (V, R, I) and near-IR bands (H)}
\begin{tabular}{ccccccccccc}
\tableline\tableline
ID & $\alpha_{2000} (^{hms})$ & $\delta_{2000} (\degr\arcmin\arcsec)$ & $P_{V}(\%)$ & $\theta_{V}(^{\circ})$ & $P_{R}(\%)$ & $\theta_{R}(^{\circ})$ & $P_{I}(\%)$ & $\theta_{I}(^{\circ})$ & $P_{H}(\%)$ & $\theta_{H}(^{\circ})$  \\ \hline
1    & 18  8 35.44 & -24 17 34.4  &$  *        $ &$  *          $ & $ 7.51(3.27)$ &$ 14.6(11.4)$ & $  *        $ &$  *       $ & $  *        $& $  *       $\\
2    & 18  8 35.59 & -23 52  5.6  &$  *        $ &$  *          $ & $ 3.62(1.38)$ &$ 86.5(10.2)$ & $  *        $ &$  *       $ & $  *        $& $  *       $\\
3    & 18  8 35.64 & -24 10 42.9  &$  *        $ &$  *          $ & $16.84(4.54)$ &$ 18.5(7.5) $ & $  *        $ &$  *       $ & $  *        $& $  *       $\\
1116 & 18  9 31.42 & -24 02 43.3  &$ 5.77(0.40)$ &$ 29.6(3.2)   $ & $ 4.80(0.23)$ &$ 22.0(1.5) $ & $ 3.78(0.59)$ &$ 38.5(4.8)$ & $  *        $& $  *       $\\
1188 & 18  9 33.54 & -24 07 18.5  &$  *        $ &$  *          $ & $  *        $ &$  *        $ & $  *        $ &$  *       $ & $ 2.28(0.34)$& $175.0(7.3)$\\
1295 & 18  9 35.73 & -24 08 07.2  &$ 5.59(1.21)$ &$  3.6(6.6)   $ & $ 3.15(0.69)$ &$177.9(6.2) $ & $ 3.32(0.42)$ &$173.4(4.0)$ & $ 1.35(0.29)$& $178.1(8.5)$\\
1399 & 18  9 37.56 & -24 08 15.0  &$ 3.64(0.48)$ &$172.1(4.5)   $ & $ 3.82(0.21)$ &$173.9(1.7) $ & $ 3.54(0.13)$ &$169.8(2.0)$ & $ 0.96(0.07)$& $163.0(6.3)$\\
1502 & 18  9 39.04 & -24 02 16.2  &$ 4.69(1.35)$ &$ 47.9(8.3)   $ & $ 4.42(0.92)$ &$ 35.8(5.9) $ & $ 4.25(0.71)$ &$ 31.9(5.1)$ & $ 1.33(0.50)$& $ 35.1(11.7)$\\
1751 & 18  9 42.67 & -24 10 41.5  &$ 1.52(0.59)$ &$101.5(10.7)  $ & $ 0.61(0.32)$ &$120.5(13.4)$ & $ 0.87(0.31)$ &$ 90.5(9.9)$ & $  *        $& $  *        $\\
\hline
\end{tabular}
\tablecomments{The table shows only some entries from the entire data set, in order to 
exhibit its content and organization scheme. Columns respectively represent 
the star's identifier number (ID), the equatorial coordinates ($\alpha,\delta$), 
together with the polarization degree and angle ($P$ and $\theta$) associated 
to each spectral band (V, R, I, and H). Values marked by * represent undetected 
sources or objects excluded from the sample. Numbers inside parenthesis are 
the computed uncertainties.
}
\end{table*}

Figure \ref{obsfields_sh229} shows a R-band DSS image of Sh 2-29, used to indicate the mapped
fields from each band and telescope, as detailed in the caption above the image.
The BDS2003 2 embedded cluster is at the center, surrounded by the arc-shaped shell.
R-band observations provide a large spatial coverage, consisting of 11 fields obtained with the OPD-0.6\,m 
telescope (solid blue line), summed to an additional field focused on the central cluster observed 
at the CTIO-0.9\,m telescope (dashed blue line). V and I bands were obtained for only one field centered 
on the embedded cluster, using the OPD-0.6\,m telescope (dot-dashed lines, respectively green and yellow).
Near-IR H band observations using the OPD-1.6\,m telescope are shown by the 5 smaller rectangular areas 
(solid red line), covering the central cluster, as well as some nearby structures. Slight variations 
in the H-band field size is a consequence of small imperfections in the jittering process.

Figure \ref{obsfields_sh229} shows that several superposition areas exist between adjacent fields, 
particularly for R-band observations. Such property was used to test the data reproducibility, 
since within these areas the same stars were observed more than once. In some cases, 
particularly near the central cluster, stars were independently observed up to 5 times. 
Hence, $P_{R}(1) \times P_{R}(2)$ and $\theta_{R}(1) \times \theta_{R}(2)$ diagrams were
built, as shown in Figure \ref{comparison_pol_sh229}, by comparing each observation pair
related to the same object. Taking into account the error bars, a good correlation is 
noted in most cases, generating a well-defined band of points around the equality line.
A larger disparity is present toward some objects in the polarization angle comparison, 
mainly related to cases where polarization degree is low, and therefore $\theta$ is not 
a well-defined quantity. To each repeated observation, the chosen value to be used on the 
following analysis was the one with larger $P/\sigma_{P}$ (best quality).
Although there is no evidence of the existence of intrinsic polarization (as will be discussed
in the next Sections), some differences may not be explained from the analysis of the error
bars, what perhaps could be related to an intrinsic time-dependent variation of polarization, 
which is common to be found specially toward young stars with circunstellar disks 
\citep[see for example, ][]{pereyra2009}.

In order to demonstrate the content and organization of the polarization values set, 
a portion of the collected data is shown in Table \ref{tab_pol_sh229}. 
The total number of detected stars in at least one of the four observed spectral bands 
sums up to 5254 objects.

Complementing the information provided by the polarimetric sample, a dust extinction map ($A_{V}$) 
was constructed from 2MASS point source catalog (PSC) JHK photometry 
data and the NICEST algorithm \citep{lombardi2009}. Sources with infrared excess, mostly candidate 
young stellar objects (YSO's) with dusty envelopes were removed from the catalog, as they have 
intrinsically red colors. The NICEST method estimates extinction by comparing the colors of 
background stars reddened by dust, with the intrinsic colors of extinction-free sources from a 
nearby control field. The map is constructed by spatial smoothing of individual extinction 
measurements, using the weighted mean of values as a function of position. NICEST combines 
Gaussian weighting and a bias correction for the decrease of the number of sources available as 
a function of extinction. We used a Gaussian width of 60 arc-second (equivalent to a beam 
resolution for the map), and Nyquist sampling. Contours exhibited in Figure 
\ref{ism_sh229} were derived from this map.

\section{Results}
\label{s:resultssh229}

\subsection{Linear polarization mapping in the R optical band}

   \begin{figure*}[!t]
   \centering
   \includegraphics[width=\textwidth]{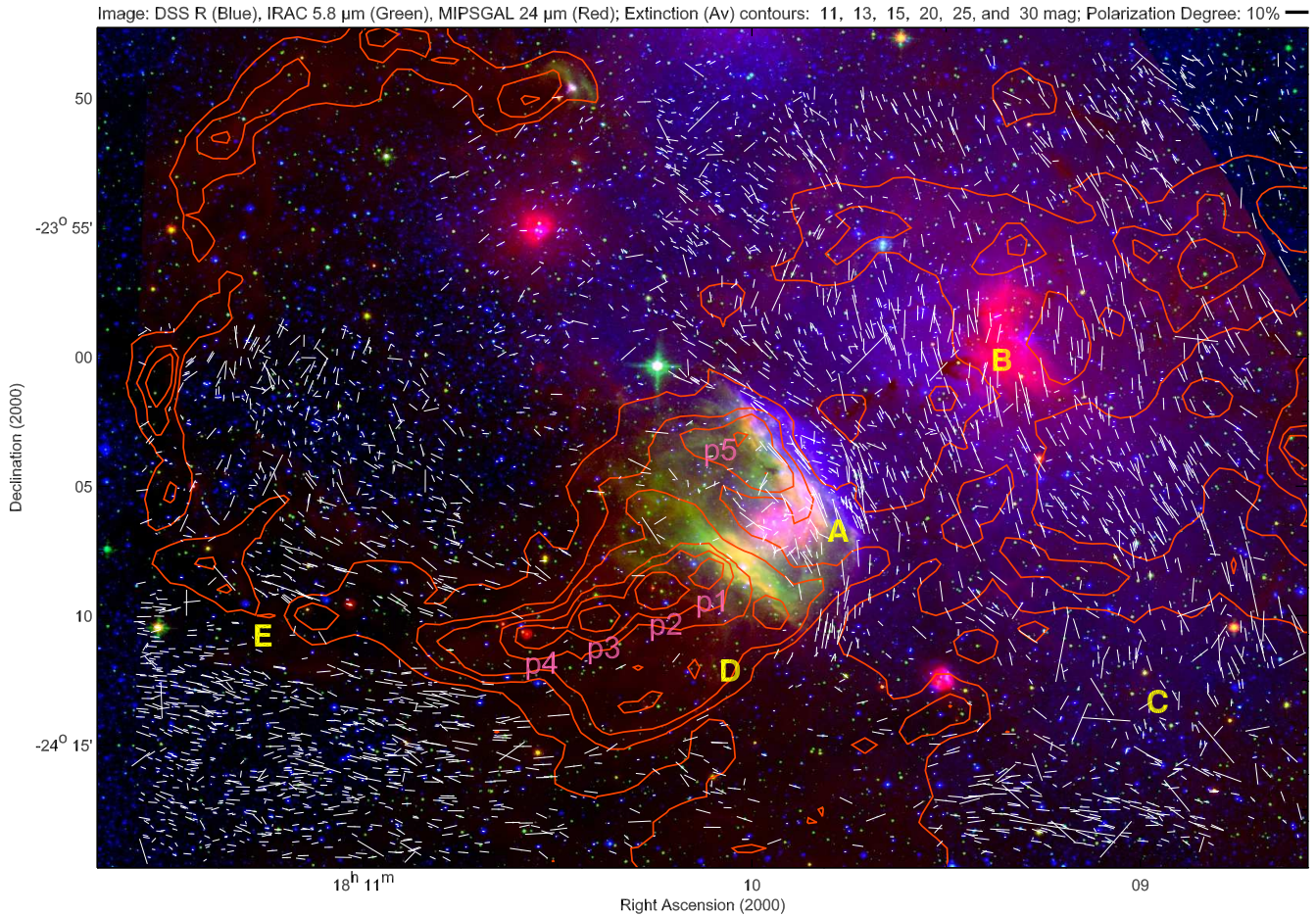} \\
      \caption{R band polarimetric mapping of Sh 2-29. The combined image is composed by 
               the R band (DSS, blue), as well as by the 5.8 and 24$\,\mu$m bands from 
               {\it Spitzer} (respectively green and red). The size of each vector is proportional to 
               $P$ (a $P=10\%$ vector is shown at the upper right). Labels from A to E 
               are used to highlight some of the main features inferred from the polarization vectors'
               orientation pattern, as discussed in Section \ref{s:polmap_qual_analysis}.
               Indicators p1 to p5 show the location of dense cloud structures around the central cavity, 
               presenting peak extinction values between $A_{V}=20$ and $A_{V}>35$ 
               (see Section \ref{s:fragsmall_sh229}).
              }
         \label{polRmap_sh229}
   \end{figure*}

%

   \begin{figure*}[!t]
   \centering
   \includegraphics[width=\textwidth]{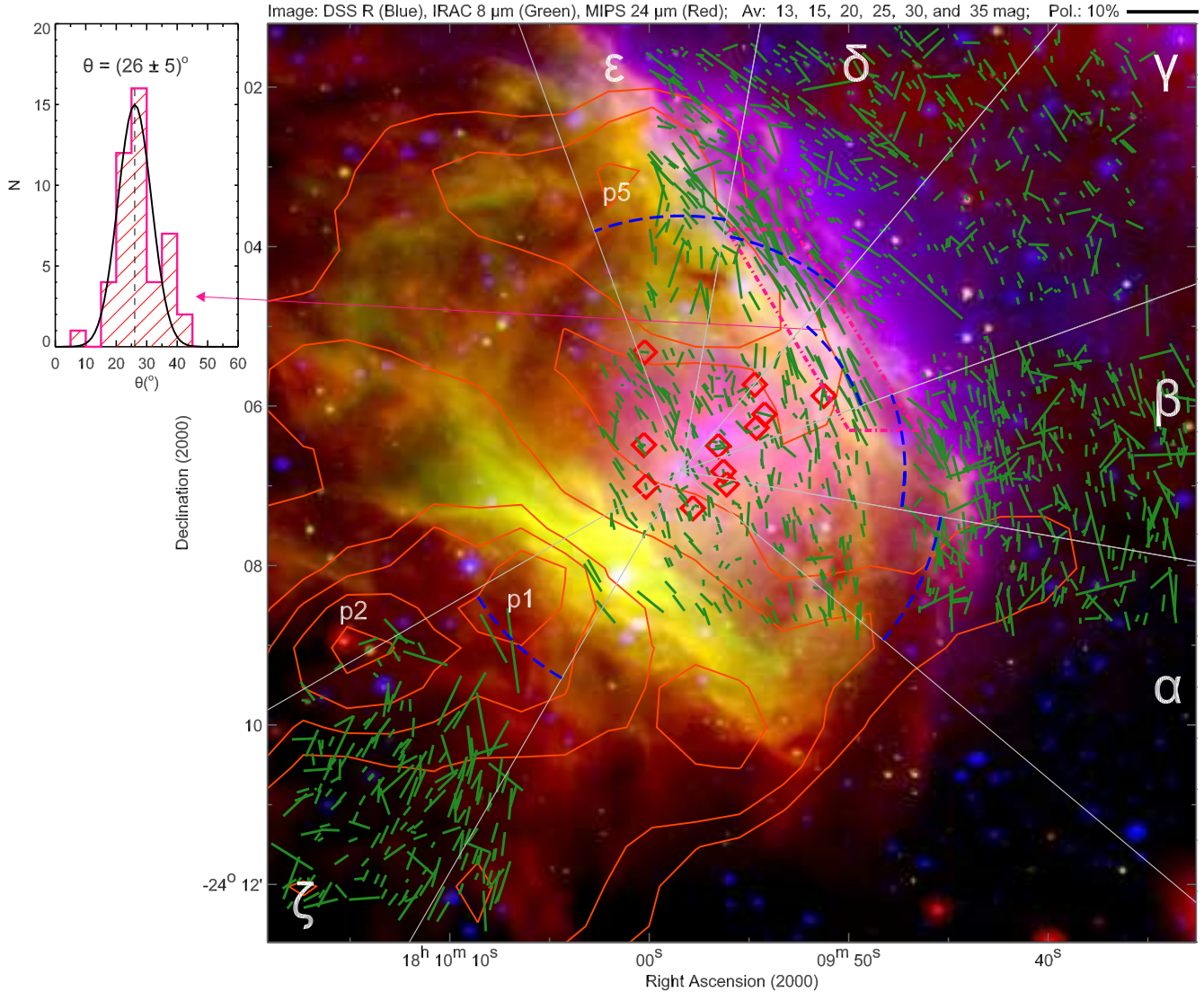} \\
      \caption{Near-IR (H band) polarimetric mapping toward the Sh 2-29 central interstellar cavity. 
               A RGB combined image of this area is used, corresponding to the R-DSS band (blue), 
               as well as the $8$ and $24 ~\mu$m bands from {\it Spitzer} (respectively green and red).
               The radial strips labeled from $\alpha$ to $\zeta$ are used  to study polarization degree as a function of
               radius (Section \ref{s:polhmap_pilling}), with the blue dashed lines indicating the positions
               were a rise in polarization occur.
               Red diamonds mark the positions of those stars presenting H$\alpha$ emission \citep{ogura2002}
               that were detected in the polarimetric survey.
               The dot-dashed pink polygon indicates the area which was used 
               to apply the Chandrasekhar-Fermi method (see Section \ref{s:chandfermish229}). The 
               polarization angle histogram shown at the left is used in this same analysis
               in order to derive the angle dispersion within this area.
              }
         \label{polHmap_sh229}
   \end{figure*}
%

   \begin{figure}
   \centering
   \includegraphics[width=0.48\textwidth]{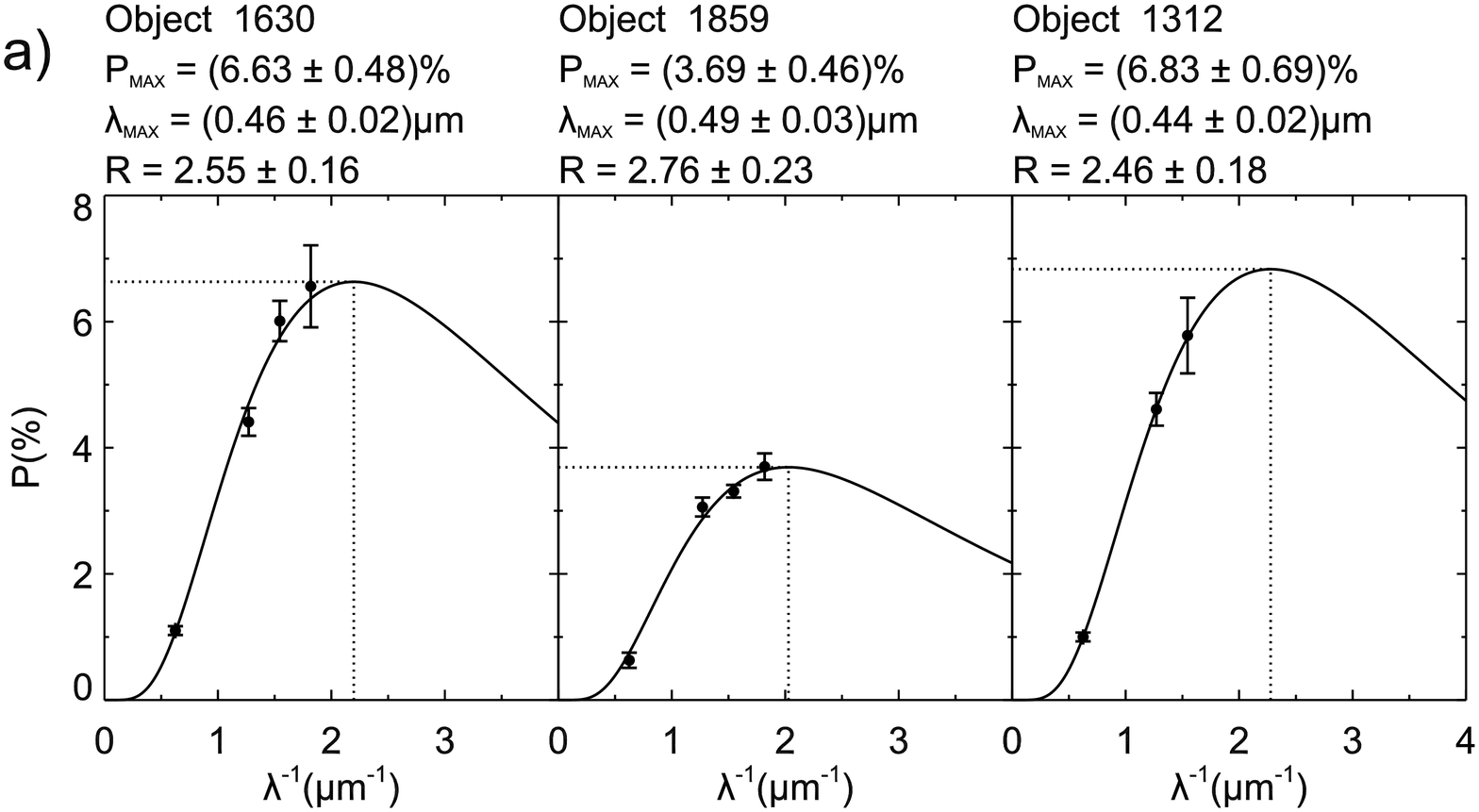} \\
   \includegraphics[width=0.48\textwidth]{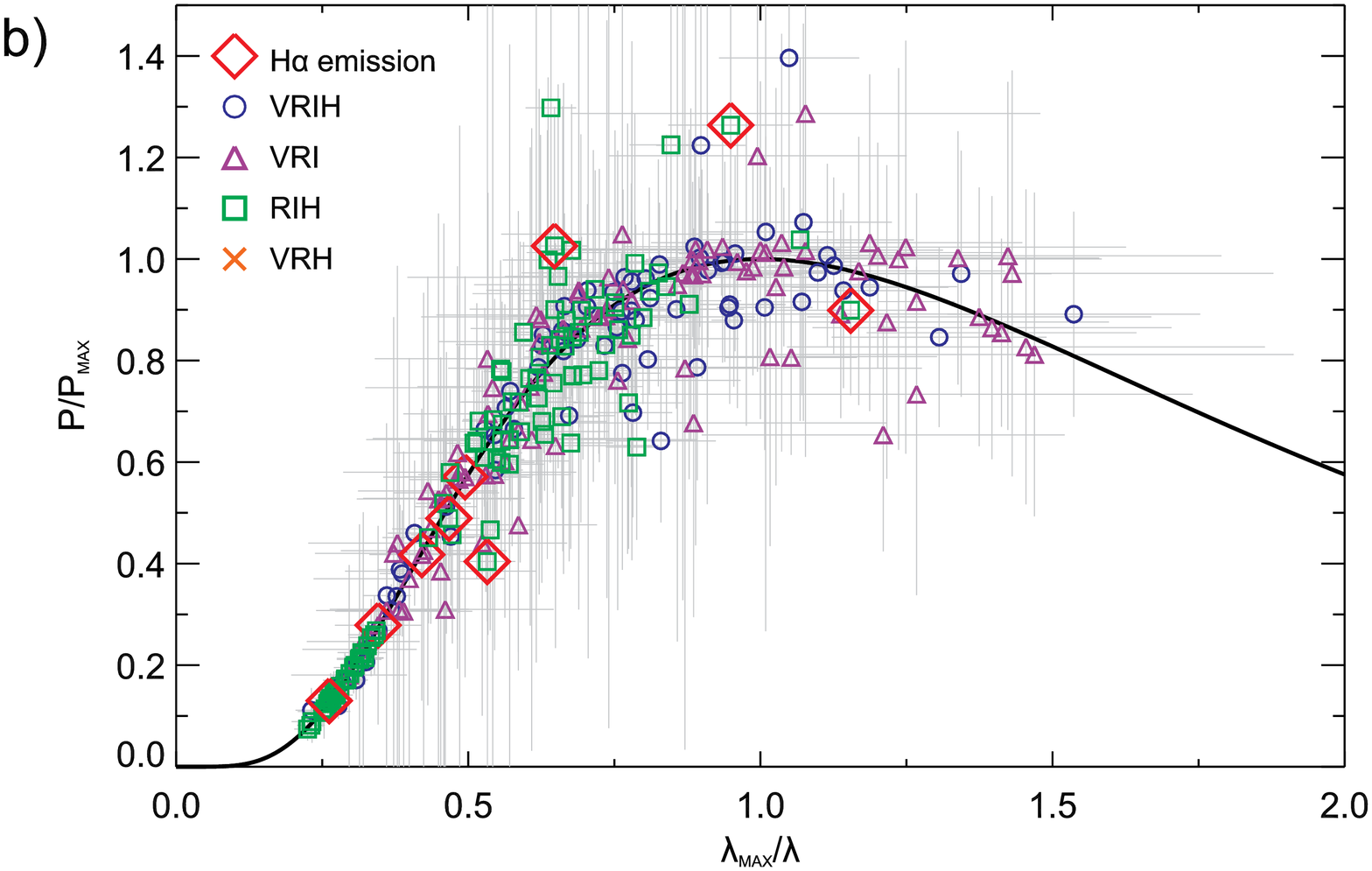} \\
   \includegraphics[width=0.48\textwidth]{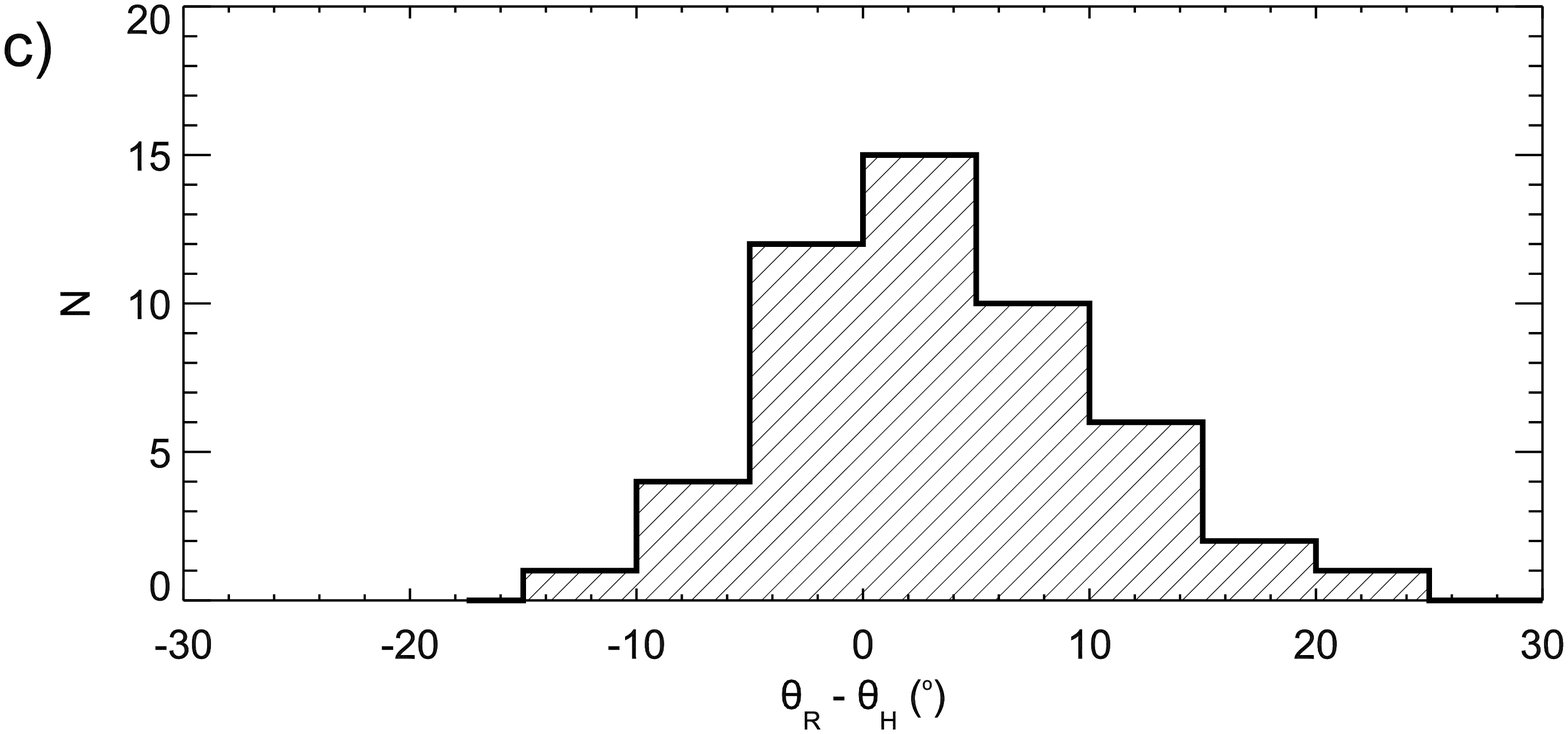} \\
      \caption{Fits of the Serkowski relation for several objects toward Sh 2-29.
         Part {\it a} shows three examples of $P(\%) \times \lambda^{-1}
         (\mu$m$^{-1})$ diagrams, where each fitted values of $P_{max}$ and
         $\lambda_{max}$ are listed at the top, as well as the total-to-selective
         extinction $R$. A $P/P_{max} \times \lambda_{max}/\lambda$ diagram
         is shown in part {\it b}, which has been built with all available polarimetric
         data for the 87 objects from the sample. The Serkowski curve is represented by
         the dark solid line, while different colored symbols denote the sets of spectral
         bands used at each fitting (as detailed by the caption inside the diagram).
         Data points marked by red diamonds indicate polarization values for 3 stars with
         H$\alpha$ emission \citep{ogura2002}. Part {\it c} exhibits a histogram of 
         $\theta_{R}-\theta_{H}$ to show the correlation in polarization angles between bands 
         R and H for all objects used in the Serkowski fits with detections in these two bands.
              }
         \label{serkfits_sh229}
\end{figure}

Figure \ref{polRmap_sh229} shows the R-band linear polarization vectors superposed to a combined
false-color image of R (DSS, blue), 5.8 and 24$\,\mu$m bands ({\it Spitzer}, respectively green and red).
Polarization degree $P$ is linearly related to the size of each vector.

Assuming that the interstellar dust grains' orientation is predominantly determined by the 
typical alignment mechanisms \citep{davis1951}, then the polarization direction is parallel 
to the sky-projected component of magnetic field lines. However, it is important 
to emphasize that the observed polarization values are not a result of a simple 
line-of-sight additive effect, and actually depend on the distance to each source and
on the variations of the interstellar medium properties along its direction. For instance,
if a star is located behind the star-forming region, its polarimetric parameters 
are represented by a combined effect of the Sh 2-29 interstellar medium 
together with the foreground dust component. Such properties will be further explored 
in Section \ref{s:forepol_sh229}.

Furthermore, it is possible that young stellar objects contribute with an 
intrinsic polarization component, which may arise, for example, if these sources present 
circumstellar disks with an asymmetrical sky projection.
This effect would probably be mostly important closer to regions with greater 
star-forming activity, such as the interior of the central interstellar cavity, 
where the embedded cluster BDS2003 2 is located. Several young stars with 
H$\alpha$ emission are found in this area, as represented by the red diamonds on Figure \ref{ism_sh229}
\citep{ogura2002}. Since most of these objects are normally too obscured, 
only a few were detected on the optical observations. 
Besides, the global orientation pattern of polarization vectors exhibit a high 
local correlation within specific areas. Such effect may only be explained due to the 
predominance of the polarization effect due to the interstellar medium. 
Therefore, the importance of a possible intrinsic polarimetric component is probably 
not statistically significant in this analysis (however, it may be important
on the analysis of the H band polarimetric properties near the central cavity, 
as will be discussed in Section \ref{s:polhmap_sh229}).

\subsection{Linear polarization mapping in the H band}
\label{s:polhmap_sh229}

Figure \ref{polHmap_sh229} shows the H-band polarimetric mapping, which covers a 
smaller area focused on the central cavity and nearby surroundings. The image consists of a RGB 
combination of the R-DSS (blue) and {\it Spitzer}'s $8$ and $24 ~\mu$m bands 
(respectively green and red).
It is evident that within the H-band survey area, a much larger number of stars were
detected, as compared to the same area at the optical observations, since interstellar
extinction is lower at near-IR bands.

At the near-IR observations, several stars with H$\alpha$ emission were detected,
as shown by red diamonds in Figure \ref{polHmap_sh229} \citep{ogura2002}.
Therefore, these are potential targets that may show some intrinsic polarization component.
In order to test this hypothesis, we compare its polarimetric parameters (degree and angle) 
with nearby objects that do not show H$\alpha$ emission. Any significant diverging trend could be 
and indication that some level of intrinsic polarization arising from the circunstellar disk
is in fact superposed to the interstellar component. 
This comparison shows that objects with H$\alpha$ emission have polarization degree values 
roughly distributed between $P_{H} = 0.5\%$ and $2.3\%$, which is similar to the levels shown by 
other nearby stars located inside the central cavity. Likewise, it is also difficult 
to distinguish any significant difference in the distribution of polarization angles for stars 
with and without H$\alpha$ emission.
These evidences suggest that, similarly to the R-band survey, it is not possible 
to infer any intrinsic polarimetric component that could be statistically significant,
therefore interfering with the analysis of interstellar magnetic field lines. As will be shown 
in Section \ref{s:polserk_sh229}, an analysis of linear polarization in multiple spectral 
bands may provide further indications to support this hypothesis. 

\subsection{Wavelength dependence of polarization and fits of the Serkowski relation}
\label{s:polserk_sh229}

In order to study interstellar grains' features toward the H{\sc ii} region's central 
cavity and close surroundings, we have used the multi-band polarimetric data to compute
fits of the Serkowski relation \citep{serkowski1975}, which may be expressed as: 

\begin{equation}
P_{\lambda}=P_{max}\exp{\left[-K\ln^{2}{\left(\frac{\lambda_{max}}{\lambda}\right)}\right]}
\label{e:serkowski_rel}
\end{equation}

\noindent Such empirical equation represents the dependence of polarization degree $P$ with 
wavelength $\lambda$, where $P_{max}$ and $\lambda_{max}$ respectively denote the maximum polarization 
level and the wavelength where this maximum value is reached. The $K$ parameter typically 
assumes the value $1.15$ \citep{serkowski1975,codina1976}.

By combining the polarimetric data in all available spectral bands (V, R, I, and H), we have
chosen objects observed at least in three different wavelengths,
and applied the fitting method proposed
by \citet{coyne1974}, which basically consists of a linearization of Equation \ref{e:serkowski_rel}.
We have also restricted the sample to those objects presenting $P/\sigma_{P}>3$ on all
available bands, resulting in 87 stars satisfying these criteria. Three examples of
fittings of the Serkowski relation ($P \times \lambda^{-1}$ diagrams)
are shown in Figure \ref{serkfits_sh229}a, together with the respective values of
$P_{max}$ and $\lambda_{max}$.

In each case, we have also
computed the parameter $R$, which represents the total-to-selective extinction ratio
($R=A_{V}/E_{B-V}$) with values shown in Figure \ref{serkfits_sh229}a for each example.
$R$ is directly linked to
$\lambda_{max}$ through the relation $R = (5.6\pm 0.3)\lambda_{max}$
\citep{serkowski1975,whittet1978}, and therefore may be easily obtained from this method.

Figure \ref{serkfits_sh229}b shows a $P/P_{max} \times \lambda_{max}/\lambda$ diagram, 
where we have used the polarization degree values
at each available spectral band for all 87 objects. Different symbols and colors are
related to the sets of spectral bands used at each fitting, as detailed by the caption
inside the diagram. Among these objects, 3 present H$\alpha$ emission
\citep[red diamonds, ][]{ogura2002}, and 82 are available at the 2MASS catalogue
(this correlation will be essential to the analysis shown in Section \ref{s:variationsR_sh229}).

Data points clearly define a thick band around the Serkowski curve in this diagram.
Although some scattering is obviously present, this is probably mainly a
consequence of the natural data uncertainties, as may be inferred by the
analysis of the error bars. Greater scattering and uncertainty levels occur for higher
$\lambda_{max}/\lambda$ values, and therefore this is mainly
related to the V and R spectral bands. This is an expected trend, since many of these stars
are highly embedded, resulting in lower signal-to-noise ratios at higher 
frequency detections, as compared to the near-IR band.

By studying the data distribution for the 3 stars with
H$\alpha$ emission (Figure \ref{serkfits_sh229}b), we notice that although some points present a significant
scattering when compared to the position of the curve, it is probably generated by the natural
scattering also detected for other stars, mainly in the V and R bands data. Therefore, we may
conclude that although some intrinsic polarization level may exist for objects
with indications of the presence of circumstellar disks, such contribution is probably too low when compared
to the prevailing dichroic effect due to magnetically aligned interstellar dust grains.

Figure \ref{serkfits_sh229}c shows a histogram of $\theta_{R}-\theta_{H}$ which have 
been constructed from all stars used in the Serkowski fits that presented detections in both 
R and H bands. Although there is a large concentration of stars with $\theta_{R}-\theta_{H}$
close to zero (indicating a correlation between these two quantities), there is clearly a 
spread in this distribution, ranging from $-15\degr$ to $\sim 25\degr$. Also, the peak of this 
distribution seems slightly displaced toward positive values. These features are expected, since 
a polarized stellar light beam that traverses different interstellar layers, presenting 
different grain properties and alignment conditions, should be affected by a rotation effect of 
the polarization angle with respect to wavelength \citep{gehrels1965,coyne1974,messinger1997}. 
Given the filamentary nature of this central area, and also the 
fact that there is a foreground polarizing dust component, the distribution's spread and displacement 
may be explained by the rotation effect.

\section{Analysis and discussion}
\label{s:analysisdiscussionsh229}

\subsection{Qualitative analysis of magnetic field lines' morphology}
\label{s:polmap_qual_analysis}

Several conspicuous features related to the general morphology of magnetic field lines may be 
immediately noticed through an overview analysis of the polarimetric mappings from 
Figures \ref{polRmap_sh229} and \ref{polHmap_sh229}. In this Section, these properties 
are discussed in detail, and some hypothesis related to the generation and evolution of these structures 
are raised. Labels from A to E are shown in Figure \ref{polRmap_sh229}
to highlight the positions of these characteristics. 

Initially, the most notable 
feature, indicated by A, corresponds to a large arc-shaped orientation of 
polarization vectors, which are curved around the central cavity, mostly following 
the same morphology of the extended H$\alpha$ emission seen in the R-band image. 
The similarity in shape between the magnetic field lines and this interstellar structure
(which is related to the ionized gas) suggests a strong correlation among these features.
Moreover, such arc-shaped characteristic, enfolding and following the cavity's 
borders, is consistent with a possible expansion of this central volume, affecting the properties of 
the surrounding regions. 

Some properties of the cavity's expansion may be inferred by studying its morphological 
features mainly through {\it Spitzer}'s mid-IR images, particularly the $5.8\,\mu$m band
(green color in Figure \ref{polRmap_sh229}). The cavity is defined by two thick 
interstellar shells -- or slabs --  which are approximately parallel to each other, and surround the 
stellar cluster located at its center. One shell is located toward the Northwest of the cluster 
(above it and to the right),  while the other is located toward the Southeastern direction
(below and to the left). By studying the $A_{V}$ contours, one may notice that above these 
shells, interstellar extinction is higher ($A_{V} \gtrsim 13$ mag, therefore indicating a higher dust and gas density), 
as compared to the nearby areas. These evidences suggest that due to the expansion of the ionized
gas inside the cavity, interstellar material has been swept out from the center (where $A_V$
is lower), being driven toward its borders, and consequently pilling up in this region.
Since magnetic fields are probably coupled to the interstellar gas, its shape
may have been affected during this process, possibly resulting in a pile-up effect of
field lines along the cavity's borders. Such effect will be further analysed in Section
\ref{s:polhmap_pilling}.

The same alignment of polarization vectors with the cavity's Northwest border is observed 
at the H-band map (Figure \ref{polHmap_sh229}, especially toward the dot-dashed pink polygon).
Furthermore, by analysing the vector's sizes, we notice that polarization degree toward such
border area seems higher on average, as compared to vectors located inside the cavity. 
Such feature seems to indicate a spatial dependence of polarization degree, especially when the 
region is analysed from inside-out (this property is further explored in detail in Section 
\ref{s:polhmap_pilling}). Apart from these trending features, the H-band mapping shows 
a large dispersion of polarization vectors' orientation (the same dispersion is observed toward the 
central cavity at the optical mapping). Given the large number of complex 
sub-structures (such as ionizing fronts, interstellar filaments, etc.), it is probably an effect of 
the line-of-sight superposition of these features toward this smaller-scale area.

A subtle arc-shaped distortion of the field is also noticed around 
the region labelled B in Figure \ref{polRmap_sh229},
although not so evident as in region A. This region surrounds the massive star HD165921, as well as 
an associated 24\,$\mu$m emission feature, visible as a red spot.
The presence of this star
(the most massive object in the H{\sc ii} region) is probably strongly affecting its nearby 
environment, ionizing, sweeping and heating the material, consequently causing a distortion of 
magnetic field lines. In general, between regions A and B polarization vector's morphology
is consistent with a flux of magnetic field lines being compressed due to the expansion of two fronts at opposite sides 
(the central cavity and the region around HD165921). Besides, around the entire area surrounding region B, 
polarization vectors seem very well ordered and uniformly oriented. If Sh 2-29 is 
regarded as a large shell in expansion around HD165921 and its associated ionized region
\citep[as discussed in Section \ref{s:descriptionsh229}, according to the interpretation by][]{yamaguchi1999},
then a higher uniformity of magnetic field lines is indeed expected along its expanding borders, 
as shown for example by simulations of expanding H{\sc ii} regions \citep{arthur2011}.
Such hypothesis will be further explored in Section \ref{s:adfRbandpol}.

At region C, an abrupt change in vectors' mean direction occurs by comparing 
the area toward its North (vectors vertically oriented) and its South (vectors 
horizontally distributed). This region indicates the Southern border of the H{\sc ii}
region, since the H$\alpha$ extended emission from ionized gas (seen in blue at the image) 
gradually diminishes from North to South around this area. Therefore, the predominantly horizontal 
polarimetric orientation below region C is probably related to magnetic field lines outside the 
H{\sc ii} region. Around region D, a Southern portion of the Simeis 188 dark cloud is present
(as revealed by the extinction contours), resulting in a significantly lower number of polarimetric
optical detections. 

At the region around E, the cloud extends to the East in a large arc shape which
is curved toward the North, defining the H{\sc ii} region's left border.
Above E region, the vectors follow the shape of the cloud, roughly bending toward the North. 
However, a large dispersion of polarization vectors exist around this area, 
suggesting that the alignment of magnetic field lines is not very prominent
(as occurs around A and B, for example). Even though, the fact that 
field lines above E in general follow the cloud's morphology (differing from the horizontal 
trend outside the H{\sc ii} region, below E) suggest that the formation of 
this cloud has somehow shaped the morphology of magnetic fields, probably by dragging its lines
along the same process.

\subsection{Analysis of the foreground R-band polarization component}
\label{s:forepol_sh229}

Before carrying on with the analysis of magnetic field lines' morphology (using the R-band map), 
it is important to evaluate the foreground polarization component, which is intrinsically present 
for all background objects. In order to apply quantitative methods of analysis,
it is necessary to remove such contribution (see Section \ref{s:foregroundremoval}), 
since it may cause a direct influence on the vectors' orientation.

   \begin{figure}[!t]
   \centering
   \includegraphics[width=0.48\textwidth]{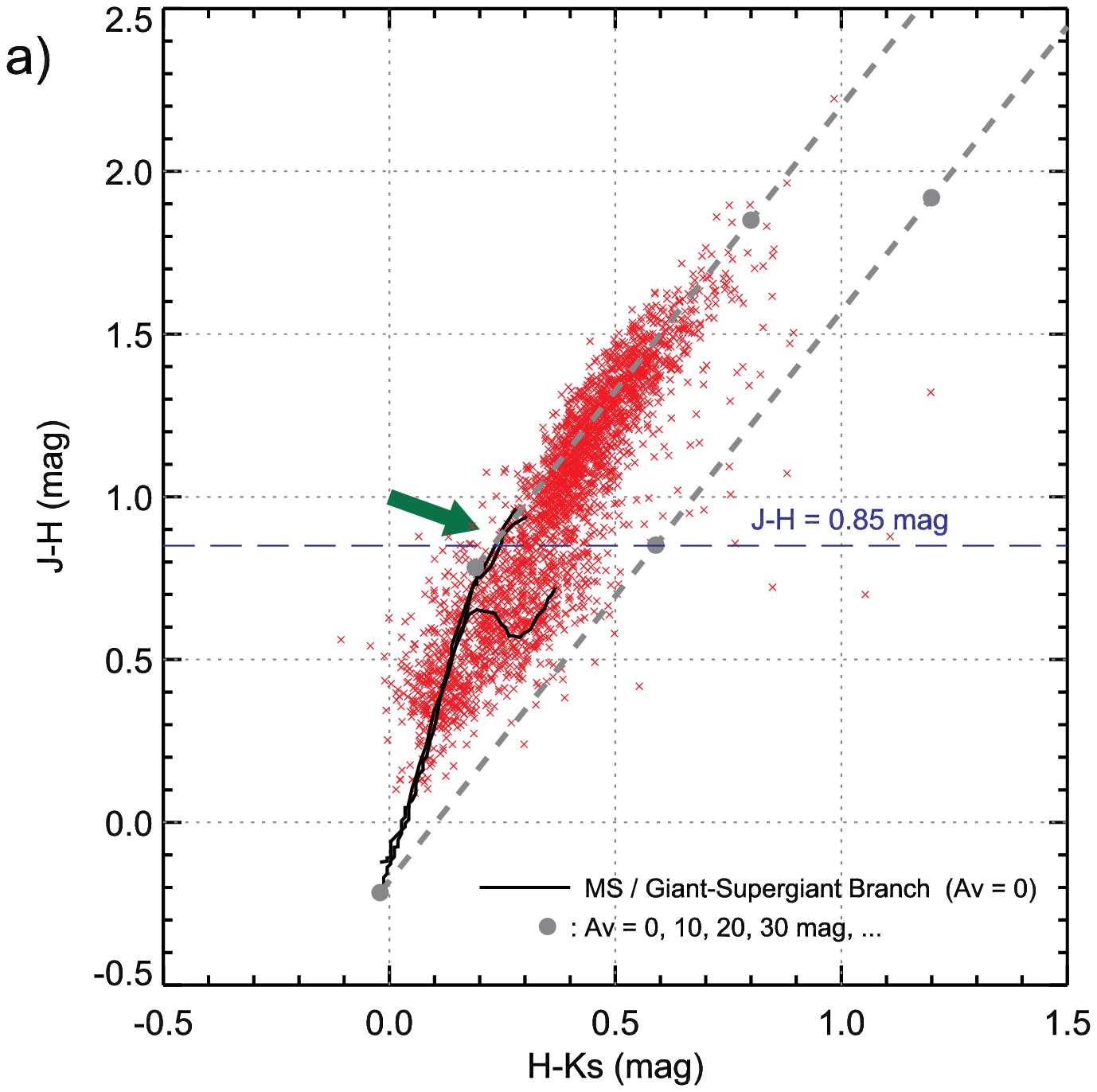} \\
   \includegraphics[width=0.48\textwidth]{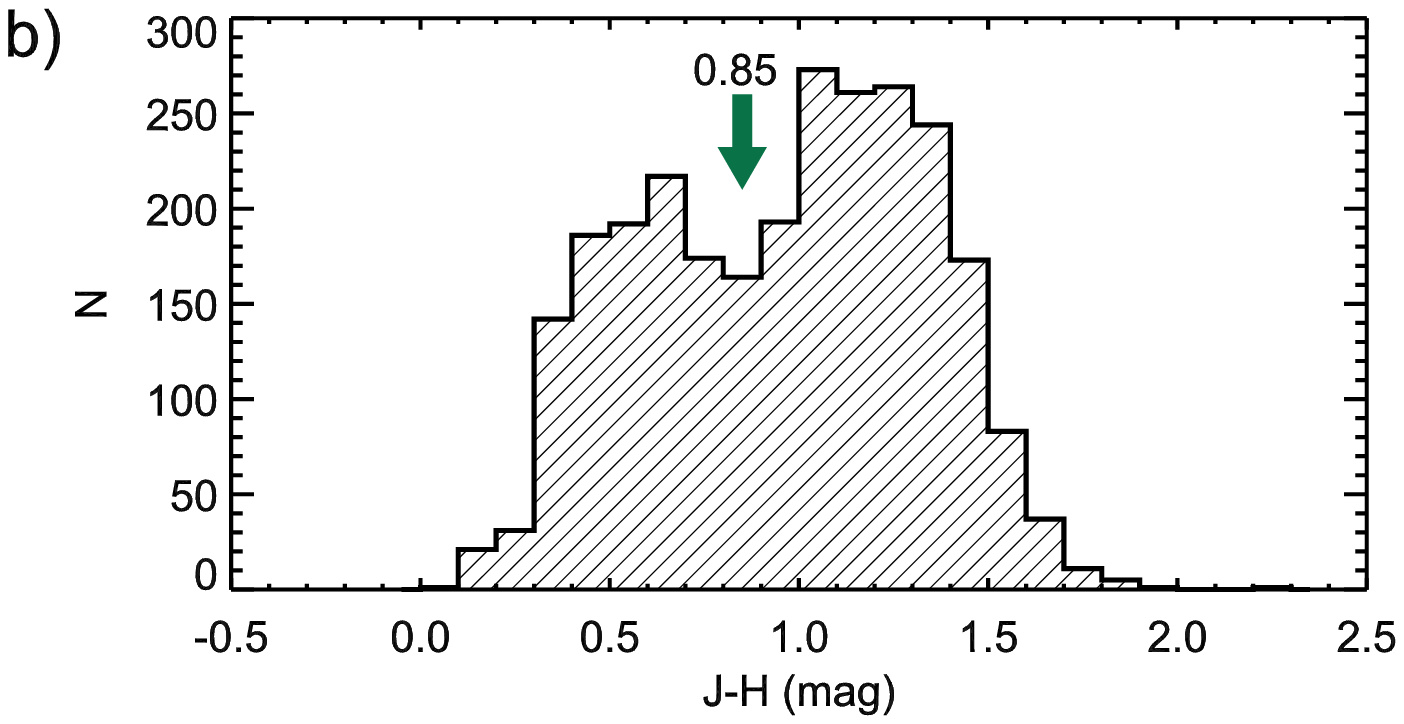} \\
   \includegraphics[width=0.48\textwidth]{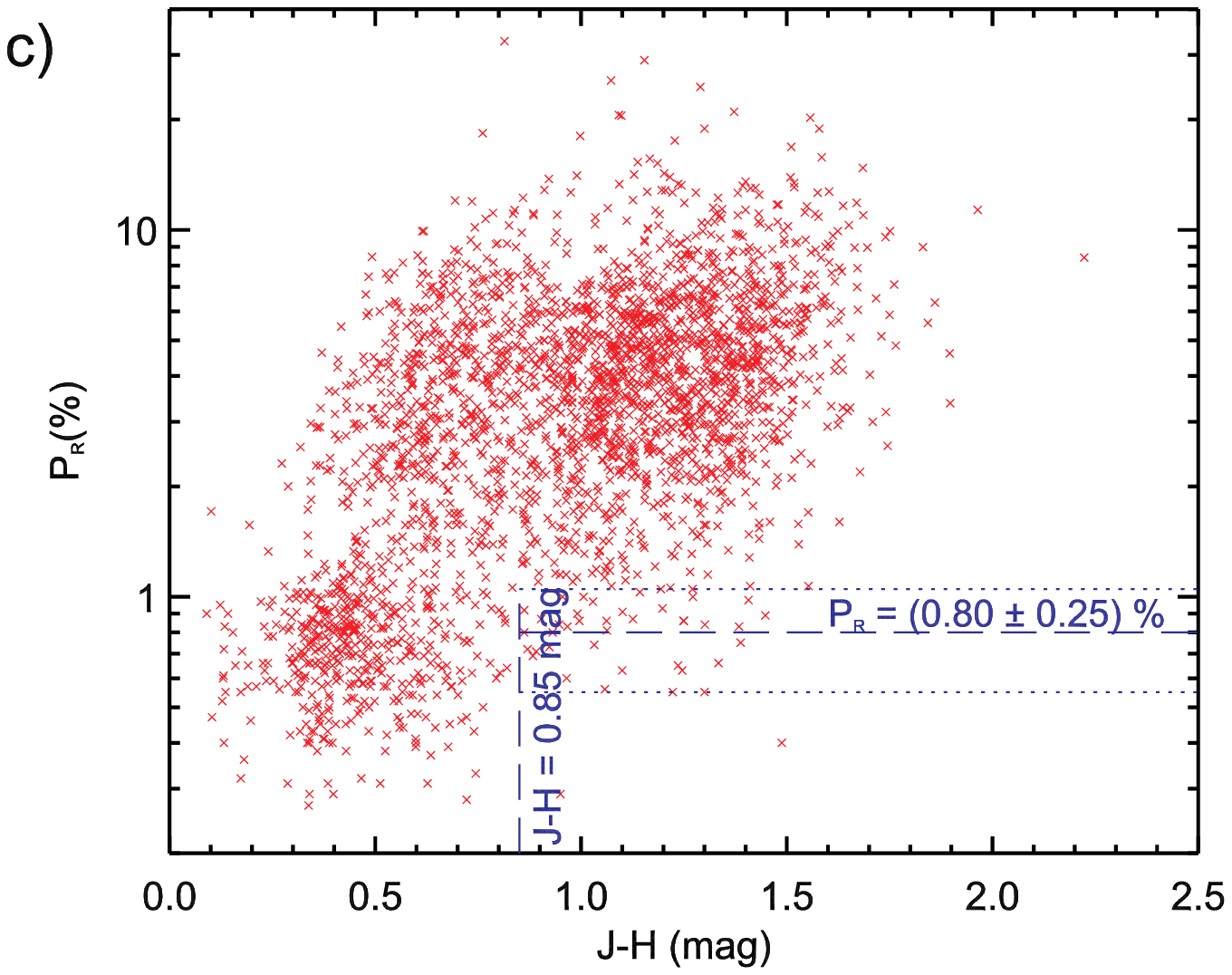} \\
      \caption{Correlation between $P_{R}$ and 2MASS near-IR data for Sh 2-29.
              Graph ({\it a}) is a color-color diagram ($J-H \times H-Ks$)
              using only stars with R-band polarimetric detections. 
              Dashed gray lines represent the reddening band \citep{rieke1985}, while
              circular dots above these lines denote positions for $A_{V} = 10, 20, 30$ mag, etc. 
              Part ({\it b}) shows a $J-H$ histogram highlighting the narrowing of points 
              at $J-H = 0.85$ (green arrow).
              Diagram ({\it c}) represents the relation between $P_{R}$ and $(J-H)$,
              using the same data set. 
              }
         \label{forepol_sh229}
   \end{figure}
%

   \begin{figure}[!t]
   \centering
   \includegraphics[width=0.48\textwidth]{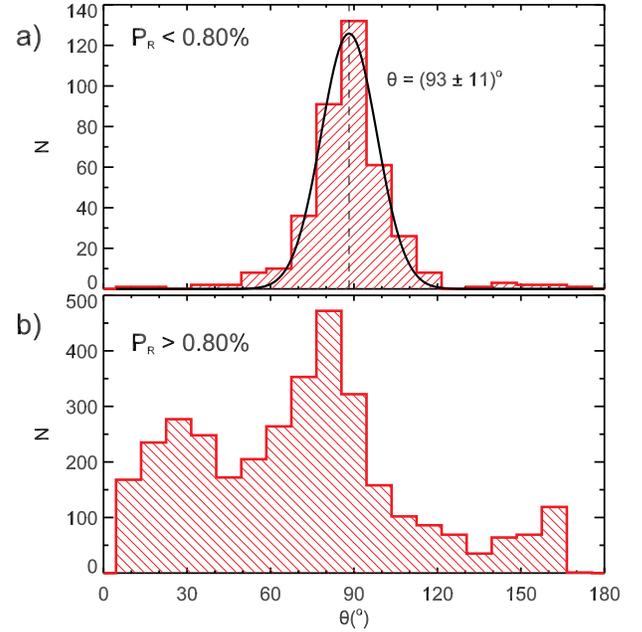} \\
      \caption{Distributions of R-band polarimetric angles, considering 
              stars with $P_{R} < 0.80\%$ (diagram {\it a}) and 
              $P_{R} > 0.80\%$ (diagram {\it b}). 
              }
         \label{forethetapol_sh229}
   \end{figure}

In order to study the extinction properties of the observed stars,
we begin by correlating the R-band polarimetric data with objects from the 2MASS catalog.
Figure \ref{forepol_sh229}a shows a $(J-H) \times (H-Ks)$ color-color diagram, made only from 
stars with R band polarimetric detections. In this diagram, dark solid lines
correspond to the un-reddened main sequence, giants and super-giants loci 
\citep{koornneef1983,carpenter2001}, while gray dashed lines represent the reddening band 
\citep[assuming the standard reddening law, ][]{rieke1985}.

A large stellar group is located around the un-reddened main sequence and giants area, 
therefore presenting lower extinction values. Moreover, several objects are spread 
along the reddening band, indicating stars with higher extinction values (and therefore 
have a higher probability of being located within or behind the star-forming region).
These two groups may be roughly separated by noting a subtle narrowing of the points'
distribution, as indicated by the green arrow, at approximately $J-H = 0.85$ mag.
Such narrowing may be clearly verified by examining the $J-H$ histogram in 
Figure \ref{forepol_sh229}b, therefore providing an approximate distinction between the lower 
and higher extinction stars.

Figure \ref{forepol_sh229}c shows a $P_{R} \times (J-H)$ diagram, built with the same 
data set used in Figure \ref{forepol_sh229}a. This type of diagram is useful to highlight
the imperfect correlation between linear polarization and interstellar extinction, an
expected trend for general polarization measurements \citep{serkowski1975}. It is obvious 
that for higher $(J-H)$ values (and therefore $A_{V}$), polarization degree average values 
are also higher. We also notice that for $J-H \gtrsim 0.85$ mag, the majority of points follow
a very clear trend: there is a minimum polarization value of approximately $0.80\%$.
Stars near this lower limit are reddened enough to present $J-H > 0.85$ suggesting 
that their location is close to or within the molecular cloud. However, their low 
polarization values indicates that they are probably not significantly affected by the 
polarizing effects due to the cloud itself, and therefore are possibly located close 
to its near edge. Thereafter, such minimum value must 
correspond to the foreground component, which is an integrated result of all interstellar 
material between the molecular cloud and the observer.
This means that all higher extinction objects (located within or behind the Sh 2-29 region), 
present this minimum polarization component value. Based on the points' dispersion 
near this minimum value, we estimate a foreground polarization degree of 
$P_{R}= 0.80\pm0.25 \%$.

In order to estimate the foreground polarization angle we use the entire R-band data set 
to build polarization angle distributions for stars with $P_{R} < 0.80\%$ (Figure \ref{forethetapol_sh229}a)
and $P_{R} > 0.80\%$ (Figure \ref{forethetapol_sh229}b). This method allows the comparison 
of polarization angles between stars averagely located in the foreground
with objects from the background. The first distribution shows $\theta_{R}$ values 
sharply concentrated around $\approx93\degr$. In fact, considering the 
relatively small area where the R-band data is distributed -- 
approximately $35' \times 45'$ -- such trend is consistent
with the fact that the foreground component should not show large angle variations.
On the other hand, the distribution of polarization angles for $P_{R} > 0.80\%$, 
shows a much larger dispersion, which is indeed expected based on the analysis of the 
large angle variations from the mapping of Figure \ref{polRmap_sh229}.
Notice that in Figure \ref{forethetapol_sh229}b there is a peak around $\theta\sim30\degr$ 
that is consistent with the polarization pattern around region B (Figure \ref{polRmap_sh229}), 
therefore clearly related to the H{\sc ii} region.
Concluding, there is a clear distinction of polarization orientation between the 
foreground population and stars affected by more distant interstellar structures.
A Gaussian fit applied to the distribution from Figure \ref{forethetapol_sh229}a
results in a foreground polarization angle of $\theta_{R}= 93\pm11 \degr$.

The foreground polarization values toward Sh 2-29 ($P_{R}$ and $\theta_{R}$)
may be compared with the foreground polarization values in the direction 
of the Lagoon Nebula, as estimated by \citet{mccall1990}. As discussed in Section
\ref{s:descriptionsh229}, this is a nearby star-forming region, located only 
$1.6\degr$ from Sh 2-29. Its foreground polarization values are  
$P= 0.81\pm0.15 $\% and $\theta= 79\pm9 \degr$ at the V optical band, which are consistent with 
our estimates, giving further support to the idea that within relatively close
areas in the sky, it is not expected that this component should suffer large 
variations. 

By converting the foreground polarization angle to galactic coordinates, we find 
$\theta_{gal}=154\degr$. Analysing large-scale Galactic polarimetric mappings 
\citep{mathewson1970,heiles2000,santos2011}, we notice that around $(l,b)\approx(7\degr,-2\degr)$, 
there is a mixture of several distinct alignments, including an orientation 
which is consistent with $\theta_{gal}=154\degr$. Toward such line-of-sight 
several sub-structures from the local interstellar medium near the Sun are known to 
exist (especially those related to the Loop I superbubble's shells), and therefore 
could be responsible for the foreground polarimetric direction
\citep[see for example,][]{santos2011}.

\subsection{Stokes parameters' removal of the R-band foreground component}
\label{s:foregroundremoval}

The R-band foreground polarimetric contribution may be removed in order to allow the 
application of quantitative methods which are dependent on polarization angles 
(see Section \ref{s:adfRbandpol}). Since a great variety of different orientations 
are present in the Sh 2-29 polarimetric map, the existence of such contribution 
causes a polarization increase or decrease, depending on whether the vector lies 
parallel or perpendicular to the foreground polarization component.

When a light beam traverses different interstellar layers, although linear polarization 
may not be considered an additive effect, its associated Stokes parameters $Q$ and $U$ 
may be successively summed. Therefore, we may consider that toward a specific object, 
the observed Stokes parameters ($Q, U$) are composed by a foreground component 
($Q_{f}, U_{f}$), added to a second component which traces Sh 2-29's magnetic 
field lines ($Q_{\mathrm{Sh2-29}}, U_{\mathrm{Sh2-29}}$): 

\begin{equation}
Q = Q_{f} + Q_{\mathrm{Sh2-29}} \ \ \ \ \mathrm{and} \ \ \ \
U = U_{f} + U_{\mathrm{Sh2-29}}
\end{equation}

The parameters associated to the foreground component were determined by 
$Q_{f}=P\cos{2\theta}$ and $U_{f}=P\sin{2\theta}$, where $P$ and $\theta$ values are those
estimated in Section \ref{s:forepol_sh229}. By subtracting these factors 
from each ($Q, U$) pair, $Q_{\mathrm{Sh2-29}}$ and $U_{\mathrm{Sh2-29}}$ were obtained,
and used to compute the corrected $P_{\mathrm{Sh2-29}}$ and $\theta_{\mathrm{Sh2-29}}$ 
values through Equation \ref{e:polparameters}.

The corrected polarization mapping is shown in Figure \ref{adfpolR_sh229}a, where 
vectors are superposed to a combined image of the R-band (DSS, blue) and $5.8\,\mu$m 
band ({\it Spitzer}, green). The visual effect is very small, as compared to 
Figure \ref{polRmap_sh229}, since the foreground polarization degree is generally 
small ($0.8\%$) relative to the typical polarization values ($\sim2-10\%$).
However, such correction is important, since the slight modification in 
individual polarization angles may result in significant changes in the statistical
analysis of the turbulence of magnetic fields (presented in the next Section).

\subsection{Turbulence of magnetic fields from the optical polarization mapping}
\label{s:adfRbandpol}

   \begin{figure*}[!t]
   \centering
   \includegraphics[width=\textwidth]{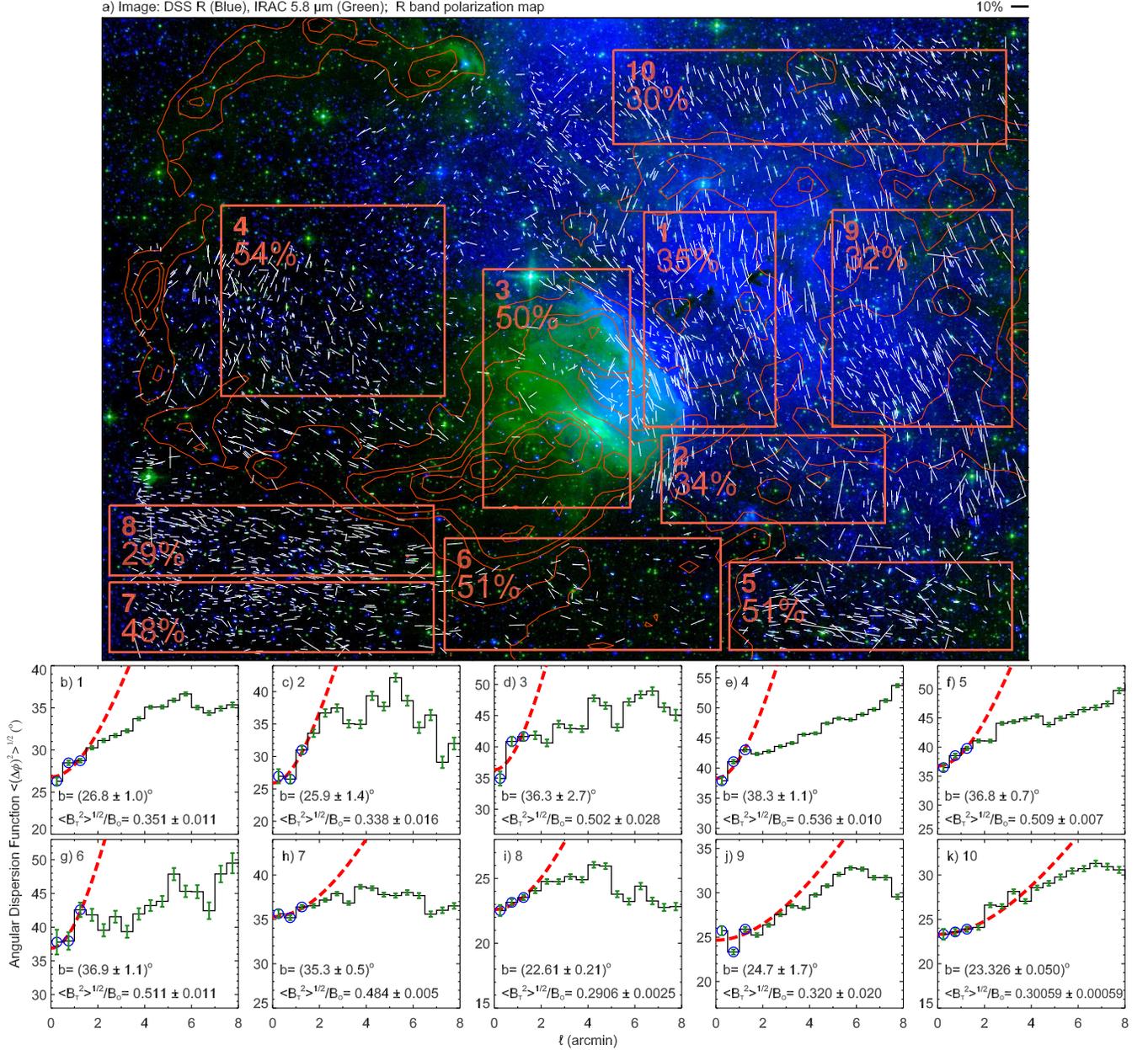} \\
      \caption{Statistical analysis of the Angle Dispersion Function (ADF) to
               different areas located toward the R-band studied region of Sh 2-29.
               Map ({\it a}) shows R-band polarimetric vectors, corrected from the 
               foreground component, superposed to a R/$5.8\,\mu$m-band combined
               image (blue/green). Different rectangular areas (1 to 10) are shown 
               over this image, and ADF diagrams ({\it b}) to ({\it k}) are respectively 
               related to these areas. Computed values of $\langle B_{t}^{2}\rangle^{1/2}/B_{0}$
               for every area are shown both inside each ADF diagram, and also inside each 
               respective area from the polarimetric map.
              }
         \label{adfpolR_sh229}
   \end{figure*}

As discussed in Section \ref{s:polmap_qual_analysis}, several hypothesis related to the interaction 
between the interstellar structures and magnetic fields toward Sh 2-29 were formulated, based solely on 
a qualitative analysis of the linear polarization mappings. Several of these hypothesis may
be tested on a quantitative manner, by studying the magnetic field's turbulence degree throughout 
the region, i.e., the ratio between its turbulent ($B_{t}$) and uniform component ($B_{0}$).

The method is based on a statistical analysis proposed by \citet{hildebrand2009}. It consists mainly 
in computing the second-order Structure Function of polarization angles 
($\langle\Delta\Phi^{2}(l)\rangle$, hereafter SF). This function may be described as 
the average value of the squared difference in polarization angles 
between two points separated by a distance $l$ \citep[see Equation (5) given by][]{falceta2008}.
Its square root is defined as the Angular Dispersion Function ($\mathrm{ADF} = \sqrt{\mathrm{SF}}$), 
and may be used to evaluate how polarization angles (and therefore magnetic field lines), 
are spatially correlated, as a function of the separation $l$. According to \citet{hildebrand2009}, 
if the turbulent and uniform fields are respectively associated to correlation lengths 
$\delta$ and $d$, then the method is valid within the range $\delta < l \ll d$. Moreover, 
the SF may be represented as an addition of a few statistically independent contributions:

\begin{equation}
\langle\Delta\Phi^{2}(l)\rangle \simeq b^{2} + m^{2}l^{2} + \sigma^{2}(l)
\label{strucfuc}
\end{equation}

\noindent The first term is related to a constant turbulent contribution $b$, 
the second term represents a smooth rise in the SF as the separation $l$ increases 
(this behaviour is considered to be linear in the ADF, with its slope denoted by 
$m$), and the third term is a roughly constant factor due to measurements uncertainties ($\sigma(l)$).
After removing the last term from each of the SF's bins, it is possible to 
compute the ADF and determine the $b$ factor (the intercept of the ADF at $l=0$)
through a simple linear fit using Equation \ref{strucfuc}. This factor is the 
essential target for this method, since it is directly related to the turbulence 
degree by:

\begin{equation}
\frac{\langle B_{t}^{2}\rangle^{1/2}}{B_{0}} = \frac{b}{\sqrt{2-b^{2}}}
\label{btb0}
\end{equation}

In order to apply this procedure to the R-band polarimetric data toward Sh 2-29 
(corrected by the foreground component) and study 
the variations of $\langle B_{t}^{2}\rangle^{1/2}/B_{0}$ throughout the region, 
different rectangular areas have been selected, labeled from 1 to 10 in Figure \ref{adfpolR_sh229}a.
The selections were based on the data distribution and on the different images' features 
observed toward each area. Also, the region's sizes were chosen such as to provide a 
sufficient number of vectors needed to perform the statistical analysis (the typical number
is between $200$ and $300$ although some regions present fewer or more vectors).
The procedure described above have been applied to polarimetric vectors inside 
each area. The respective ADF diagrams are shown in Figures from \ref{adfpolR_sh229}b to 
\ref{adfpolR_sh229}k, together with the fitted values for the $b$ parameter and the magnetic turbulence 
degree $\langle B_{t}^{2}\rangle^{1/2}/B_{0}$ (these are also shown inside each area 
of Figure \ref{adfpolR_sh229}a). In each case, $0.5'$-sized bins were used, 
and only the first three points were considered in the linear fit to Equation \ref{strucfuc}, 
in order to guarantee the validity of the condition $l \ll d$.

Regions 1, 2, 9 and 10 coincide with the area that supposedly have been affected by the 
H{\sc ii} region's expansion, and also where a higher intensity of the H$\alpha$ extended 
emission is noted (blue color in the composed image).
Particularly, region 1 corresponds to the area located among two possible fronts expanding
toward each other (the first due to the central cavity, and the second related to the massive 
object HD165921), as discussed in Section \ref{s:polmap_qual_analysis}. Comparing magnetic 
turbulence degree values within these areas, with more external regions (such as 5, 6 and 7), 
considerably lower levels are found (roughly from 30 to 35\%). 
This trend is consistent with the compression and consequent uniformity of magnetic 
field lines, due to the expansion of the H{\sc ii} region's surrounding shell, 
lending support to the previously detailed hypothesis.

Areas 5, 6 and 7 denote the H{\sc ii} region's external regions, below its Southern 
rim. This entire area is dominated by horizontally oriented vectors, differing 
significantly from the pattern observed inside the ionized volume. Furthermore, 
$\langle B_{t}^{2}\rangle^{1/2}/B_{0}$ values also diverge significantly, showing 
higher levels (between 48 and 51\%), suggesting that this area probably have 
not been affected by any ordering effect and/or dragging of magnetic field lines. 

A particularly interesting trend may be noticed by the comparison between regions 
7 and 8: although these areas are spatially close and apparently both unaffected by the
H{\sc ii} region's expansion, magnetic turbulence degree toward region 8 (closer to the dark cloud's
arc-shaped rim) is much lower ($29\%$) relative to region 7 ($48\%$) located 
immediately to the South. This may suggest that closer to the cloud, the field
lines' orientation pattern is more uniform and less random. Besides,
in Section \ref{s:polmap_qual_analysis} we have suggested that the same process which have 
generated the arc-shaped cloud's morphology, may also have induced the orientation 
of magnetic field lines parallel to the cloud (region E in Figure \ref{polRmap_sh229}). 
The correlation of these information points toward a possible explanation for the 
difference in magnetic turbulence degree between areas 7 and 8: the process that 
distorted the cloud into an arc-shape was followed by a compression and pile-up effect
of magnetic field lines behind the expanding cloud area, consequently creating 
a more uniform pattern closer to the cloud. Therefore, such interpretation 
is consistent with the fact that $\langle B_{t}^{2}\rangle^{1/2}/B_{0}$ decreases
nearer to the cloud around this region.

During the evolution of an H{\sc ii} region, it is expected that more evolved 
areas cease to expand after several million years, due to the balance between internal and external pressures, 
leading to an equilibrium state \citep{spitzer1978}. Therefore, it is not possible 
to assert that the expansion which has possibly distorted the cloud, together with 
the interstellar magnetic field lines, still exists. 
If Sh 2-29's eastern area is more evolved, this could explain the fact that 
toward area 4, a high magnetic turbulence degree is detected (54\%): 
if field lines have previously been ordered due to expansion and compression effects, 
the evolution of the system could have led to a subsequent interaction 
with interstellar gas' inherent turbulence. Therefore, such interaction
may have been responsible for rearranging and distorting its organized pattern, 
consequently increasing its random magnetic component.

Other hypothesis that could explain the relatively higher turbulence degree 
towards area 4 is related to its distance from the region's ionizing objects: 
the nearer known massive object that could ionize this region is HD166192, of spectral type B2{\sc ii}. 
It may be noted that the H$\alpha$ extended emission is relatively less intense toward this area, 
which could indicate a lower ionization level. Therefore, it is probable that the 
flux of ultraviolet ionizing photons from this star have a lower radius range, 
if compared for example, to HD165921 (spectral type O7{\sc v}+O9{\sc v}), which 
is likely responsible for the ionization and expansion effects around areas 
1, 2, 9 and 10. Consequently, it is also possible that the expansion effect 
toward area 4 is not very intense, not causing a great disturbance in the surrounding clouds, 
and therefore, not capable of producing a sufficiently pronounced ordering of 
magnetic field lines. 

Toward the central cavity (area 3), a high level of magnetic turbulence is also 
detected (50\%). Since many interstellar sub-structures and filamentary clouds
exist toward this area, it is quite possible that this value is a result of the 
superposition of differently oriented field lines related to these small scale features.

\subsection{Pilling up of magnetic field lines around the cluster's cavity}
\label{s:polhmap_pilling}

   \begin{figure}[!t]
   \centering
   \includegraphics[width=0.48\textwidth]{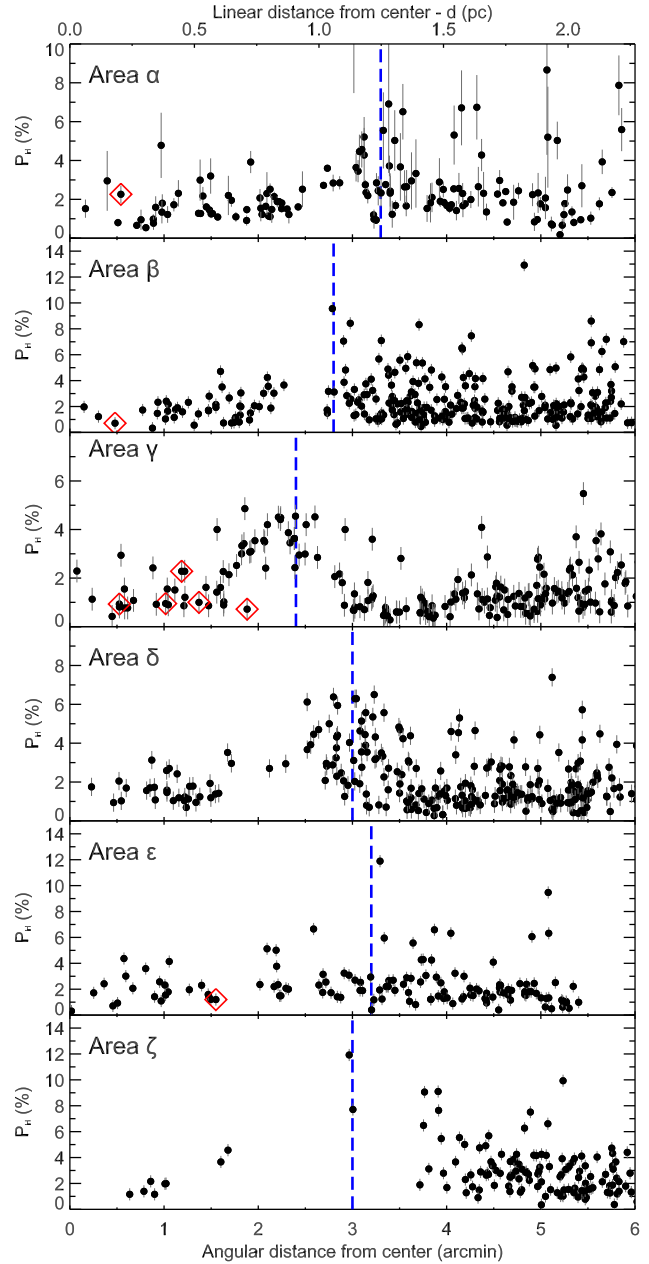} \\
      \caption{H-band polarization degree ($P_H$) as a function of the distance $r$ 
               from the center of Sh 2-29's interstellar cavity.
               Each diagram is correlated to one of the radial sections labelled as 
               $\alpha$ to $\zeta$ in Figure \ref{polHmap_sh229}. The horizontal axis from the bottom corresponds 
               to the angular radial distance (in arcmin), while the top axis represents the linear 
               distance (in parsecs, assuming a distance of $d=1.3$ kpc). The blue dashed vertical lines
               indicate the positions where we notice an increase in $P_{H}$ mean value (or at least to 
               some few objects), as compared to the levels closer to the center. Red diamonds indicate
               stars with H$\alpha$ emission.
              }
         \label{polHradial_sh229}
   \end{figure}

In order to study de behaviour of polarization degree as a function of the radius 
toward the central cavity, we have chosen radial sections in Figure \ref{polHmap_sh229} (labelled $\alpha$ to $\zeta$),
originating from the cavity's center, where the BDS2003 2 cluster is 
located. These areas were selected based on the data distribution,
providing at least $100$ vectors were found inside each sector.
$P_{H} \times r$ diagrams have been built to each section, as shown in Figure 
\ref{polHradial_sh229}.

Typically, $P_{H}$ levels between $0.5$ and $3\%$ are found inside the cavity. By analysing the 
$P_{H} \times r$ diagrams, we notice that for every radial section, a rise of the typical $P_{H}$
value occurs at a certain radius, as indicated by the blue dashed lines. Such increase
is more or less prominent, depending on the studied section. For example, regions $\gamma$ and $\delta$ show 
a sharp rise in $P_{H}$ (assuming values between $\approx2$ and $7\%$) at approximately 
$r=2 - 3$ arcmin. On the other hand, regions $\alpha$, $\beta$, $\epsilon$ and $\zeta$ show 
a less marked rise in $P_{H}$, with some sparse points assuming higher values, as compared to 
levels found inside the cavity.

In Figure \ref{polHmap_sh229} we show the same approximate positions where the polarization 
elevation occurs (blue dashed lines for each section), indicating a good general 
correlation with the location of the cavity's edge. 
We may interpret this effect as a result of the pilling up of magnetic 
field lines at these borders, which were dragged outwards due to the volume's expansion from
the action of young massive stars and ionized gas inside the region.
The result of this pile-up effect is a higher intensity of the magnetic field
at this area, which probably causes a direct influence on the efficiency of the dominant
grain alignment mechanism. Therefore, a higher value of the magnetic field is probably
responsible for a better alignment of interstellar dust grains at the cavity's borders, 
leading to higher detected values of linear polarization.

\subsection{Interstellar extinction peaks around the central cavity}
\label{s:fragsmall_sh229}

Examining the Sh 2-29 extinction map we notice the existence of some 
highly obscured interstellar extinction peaks located around the cavity's edge. These 
objects are labelled as p1 to p5 in Figure \ref{polRmap_sh229}, and present 
peak extinction values between $A_{V}=20$ and $37$ mag. Structures p1 to p4 
are approximately aligned within the dark cloud, below and to the left of
the central cavity (all presenting peak $A_{V} > 28$ mag), while p5 
(peak $A_{V} = 20.9$ mag) is located toward the North, coinciding with the 
cavity's northern edge (see Figures \ref{polRmap_sh229} and \ref{polHmap_sh229}).

\begin{table*}[!t]
\centering
\caption{\label{t:dense_clumps} Properties of dense interstellar structures around Sh 2-29's central cavity}
\begin{tabular}{cccccccccc}
\tableline\tableline
Core ID  & $\alpha_{2000} (^{hms})$ & $\delta_{2000} (\degr\arcmin\arcsec)$ & Ang$.$ Diam$.$ ($'$) & Diam$.$ (pc) & Peak $A_{V}$ & Mean $A_{V}$ & $N_{\mathrm{H_{2}}}$ (cm$^{-2}$) & $n_{\mathrm{H_{2}}}$ (cm$^{-3}$) & $M_{\mathrm{H_{2}}}$ (M$_{\odot}$) \\ \hline
p1 &  18 10 15.1  & -24 09 01.3   & $\sim$1.50  & 0.57  & 34.2  &  30.1  & $2.8 \times 10^{22}$  &  $1.6 \times 10^{4}$  &  77 \\ \hline
p2 &  18 10 06.4  & -24 08 28.4   & $\sim$1.75  & 0.66  & 36.6  &  32.5  & $3.1 \times 10^{22}$  &  $1.5 \times 10^{4}$  & 112 \\ \hline
p3 &  18 10 23.6  & -24 10 01.4   & $\sim$1.50  & 0.57  & 33.5  &  30.5  & $2.9 \times 10^{22}$  &  $1.6 \times 10^{4}$  &  77 \\ \hline 
p4 &  18 10 32.7  & -24 10 59.6   & $\sim$1.25  & 0.47  & 28.4  &  26.7  & $2.5 \times 10^{22}$  &  $1.7 \times 10^{4}$  &  46 \\ \hline 
p5 &  18 10 01.6  & -24 03 02.7   & $\sim$2.00  & 0.75  & 20.9  &  17.5  & $1.6 \times 10^{22}$  &  $0.7 \times 10^{4}$  &  76 \\ \hline 
\hline
\end{tabular}
\tablecomments{
The columns respectively represent the extinction peak identifier (see Figure \ref{polRmap_sh229}), the 
equatorial coordinates ($\alpha,\delta$), the approximate angular and linear diameter (assuming 
a distance of $1.3$ kpc), the peak and mean extinction value inside the area comprised by each structure, 
the $H_{2}$ column density (assuming the standard gas-to-dust relation based on the mean extinction, 
and considering only the presence of molecular hydrogen), the mean $H_{2}$ density and the 
$H_{2}$ Mass (assuming approximate spherical morphologies). 
}
\end{table*}

Based on each peak's extinction profile, we have estimated several physical properties, 
which are shown in Table \ref{t:dense_clumps}. An approximate angular diameter
is computed (in arcmin) and subsequently converted to linear diameter (in parsecs), 
by assuming a distance of $1.3$ kpc \citep{fich1984}. In order to estimate the average molecular 
hydrogen density ($n_{\mathrm{H_{2}}}$) and mass ($M_{\mathrm{H_{2}}}$), 
we begin by computing the mean $A_{V}$ within each peak's projected area. 
We assume that the gas-to-dust ratio between hydrogen's total column density
(i.e., neutral and molecular, $N_{\mathrm{H(total)}} = N_{\mathrm{H I + H_{2}}}$), 
and color excess $E(B-V)$ is given by de $5.8 \times 10^{21}$ particles cm$^{-2}$ mag$^{-1}$ 
\citep{jenkins1974,bohlin1978,pineda2010}. Assuming further that all hydrogen content
within such high-density volume is constituted by its molecular form, then   
$N_{\mathrm{H(total)}} = 2 N_{\mathrm{H_{2}}}$, and therefore
$N_{\mathrm{H_{2}}}/E(B-V) = 2.9 \times 10^{21}$ cm$^{-2}$ mag$^{-1}$. 
In order to convert between $E(B-V)$ and $A_{V}$, we assume the typical interstellar law
$A_{V} = 3.1 E(B-V)$. Finally, the relation between molecular gas and extinction
is $N_{\mathrm{H_{2}}}/A_{V} = 9.4 \times 10^{20}$ cm$^{-2}$ mag$^{-1}$.
By means of such relation, we have computed the $\mathrm{H_{2}}$ column density
toward each extinction peak, based on the $A_{V}$ mean values shown in Table \ref{t:dense_clumps}.
Assuming that the structure's projected diameter is similar to the line-of-sight dimension, 
the mean molecular hydrogen density ($n_{\mathrm{H_{2}}}$)  was estimated. Also, 
considering approximately spherical shapes, the $\mathrm{H_{2}}$ mass could be deduced.

Molecular hydrogen density and mass values are respectively in the range $0.7$ -- $1.7 \times10^{4}$
cm$^{-3}$ and $46$ -- $112$ solar masses. Furthermore, its dimensions vary between 
$\sim 0.40$ and $0.80$ pc. 
Such estimates suggest that these fragmentations could represent dense interstellar clumps, as compared 
for example, with studies from \citet{williams1998} and \citet{roman-zuniga2012}. Therefore, 
each of these individual structures may be considered as important targets for future small scale and high
resolution observations. 
However, such values should be regarded as rough estimates, 
in view of the number of assumptions used in the computations. For instance, 
variations in the total-to-selective extinction ratio are frequent inside molecular clouds and 
H{\sc ii} regions, typically varying between $R=2.5$ and $5.0$ (see Section \ref{s:variationsR_sh229}).
Furthermore, variations in distance and in the molecular hydrogen fraction could cause
large changes in the physical values. Also, the extinction map has a limited resolution, 
so that other gas and dust tracers at radio and sub-millimeter spectral bands could 
be used for a better characterization of these targets. 
Therefore, we estimate that variations of approximately $\approx50\%$ could occur for the 
$n_{\mathrm{H_{2}}}$ and $M_{\mathrm{H_{2}}}$ values.

It is quite remarkable, however, that by comparing the position of these dense interstellar fragmentations,
with {\it Spitzer}'s $24\,\mu$m emission (the red image component in Figures 
\ref{polRmap_sh229} and \ref{polHmap_sh229}), 
we notice the existence of some point-like sources, with a significant flux in such spectral band. 
These sources are particularly close to peaks p2 and p4, respectively at positions 
$(\alpha,\delta)_{J2000} = (18^{\mathrm h}10^{\mathrm m}15.4^{\mathrm s},-24\degr08'55.7'')$ and
$(\alpha,\delta)_{J2000} = (18^{\mathrm h}10^{\mathrm m}34.8^{\mathrm s},-24\degr10'43.1'')$.
Other mid-IR point-like sources may also be identified by the inspection of the regions close 
to dense extinction peaks p1-p4. The emission feature from this spectral region is usually linked 
to young stellar objects, presenting thermal dust emission due to the circumstellar disk. 
Therefore, this scenario is consistent with the presence of proto-stellar 
objects ongoing formation inside the dense interstellar fragmentations. 

\subsection{The relation among the interstellar extinction peaks and the magnetic 
field lines' morphology}
\label{s:fragsmall_magnetic}

A comparison between the dense extinction peak's properties and the magnetic field lines
(as revealed by Figure \ref{polHmap_sh229}) also leads to interesting results.
Among those interstellar structures indicated in Table \ref{t:dense_clumps}, p5 is the one 
with lower peak $A_{V}$ and $n_{\mathrm{H_{2}}}$. This extinction peak is located toward the cavity's
northern edge, where field lines seem to pile-up parallel to the border, therefore
roughly perpendicular to the interstellar material expansion direction.
Besides, p5 coincides with the extended structure detected at $450$ and $850\,\mu$m bands by 
\citet{morgan2008} (which is also coincident with IRAS 18068-2405 source). 
As described in Section \ref{s:descriptionsh229}, such feature was studied by \citet{urquhart2009},
who concluded that mid-IR point-like sources still does not exist within this area,
and that the interstellar density enhancement has probably been induced due to some 
external event.

These hypothesis are very consistent with our interpretation that the central interstellar
cavity is expanding, consequently leading to a pilling up of interstellar 
material along its borders, and therefore probably being responsible for the formation of 
p5 during this same process. Moreover, it is possible that the expansion direction 
perpendicular to the field lines may have contributed to detain the advancing ionization front, 
hindering a larger pilling up of interstellar material in this area 
(as compared to peaks p1-p4, for example), due to the higher magnetic pressure.
Other possibility, however, is that the differences in density between the interstellar
extinction peaks is otherwise due to simple non-uniformities in the initial surrounding medium,
before being affected by the effects of the cavity's expansion.

In contrast, toward fragmentations p1-p4 the polarization vectors' orientation suggest that 
field lines are rather perpendicular to the cavity's borders, and therefore are probably 
roughly parallel to its expanding flux. We may infer that such configuration
possibly facilitates the expansion of the cavity's southeastern rim, leading toward the 
dark cloud's densest portion (where peaks p1-p4 are located). Such expansion may have been 
responsible for generating instabilities and fragmentations, which led to the 
interstellar material contraction and collapse into highly obscured structures 
(peak $A_{V} > 30$ mag). Several models describe the triggered collapse of interstellar clumps and cores
due to externally induced shock waves, as discussed for example, by 
\citet{boss1995}, \citet{hennebelle2003,hennebelle2004}, \citet{whitworth2007} and 
\citet{andre2009}.
As previously suggested, such collapse possibly resulted in the formation of the next
generation of proto-stars with significant mid-IR emission.

\subsection{Magnetic field strength at the line compression zone}
\label{s:chandfermish229}

In Section \ref{s:polhmap_pilling} we have suggested that
the increase in polarization degree detected toward the central cavity's 
borders is a consequence of the larger magnetic field intensity,
due to the pilling up effect.
If this idea is correct, a direct estimate of the magnetic field strength in this area
should provide a higher value, as compared to typical levels observed 
at the diffuse interstellar medium.

A relatively simple procedure to estimate $B$ is the Chandrasekhar-Fermi (CF) method
\citep{chandrasekhar1953}, which is based on the idea that magnetic field irregularities 
are a natural consequence of turbulent motions from the interstellar medium. 
Such relation is obvious, since field lines are coupled to the interstellar gas, 
providing that at least a small ionization fraction exists \citep{spitzer1978,heiles2005}.
These irregularities will produce a larger dispersion in the orientation of 
polarization vectors. However, for a larger $B$ value, field lines' shape is less
susceptible to the turbulence effects. Therefore, a lower polarization angle dispersion 
results in a higher $B$ level, and vice-versa. The sky-projected field strength 
may be expressed as \citep{crutcher2004,heiles2005}:

\begin{eqnarray}
B &&= Q \sqrt{4\pi\rho}\;\frac{\delta V}{\delta\theta}\; \nonumber \\
&&\approx 9.3 \biggl(\frac{n_{\mathrm{H_{2}}}}{\mathrm{cm}^{-3}}\biggr)^{1/2}
\biggl(\frac{\Delta V}{\mathrm{km s^{-1}}}\biggr)
\biggl(\frac{\delta\theta}{1\degr}\biggr)^{-1} \; \mu \mathrm{G}
\label{e:chandfermi}
\end{eqnarray}

\noindent 
where $\delta V$ is the velocity dispersion ($\Delta V = \sqrt{8 \ln{2}}\;\delta V$ 
is the FWHM corresponding to the spectral line used in this computation),
$\rho$ is the gas density (where we assume the predominance of molecular hydrogen),
and $\delta\theta$ is the polarization angle dispersion. The parameter Q, which 
has the approximate value of $0.5$ 
\citep[according to molecular cloud simulations,][]{ostriker2001}, 
corresponds to a correction and calibration of the CF relation, in order to account for
several factors, such as line-of-sight field fluctuations,
velocity perturbations anisotropies, gas density inhomogeneities, etc
\citep{zweibel1990,myers1991}.
Nevertheless, according to the same simulations, this method is valid only for 
low dispersion values ($\delta\theta < 25\degr$).

The CF method was applied to the cavity's border area indicated 
by dot-dashed pink polygon in Figure \ref{polHmap_sh229}. The $P \times r$ diagrams 
toward this region (Figure \ref{polHradial_sh229}, sections $\gamma$ and $\delta$), 
show a sharply increase in polarization degree. The polarization angle histogram shown at the top left 
of Figure \ref{polHmap_sh229} exhibit a distribution highly concentrated 
around $\theta = 26\degr$, with a dispersion of only $\delta\theta = 5\degr$ 
(computed through the Gaussian fit). It is important to account for the measurement
uncertainties, which contribute with a fraction of this dispersion value.
The mean value of polarization angle uncertainty for vectors inside the polygon, 
is $\langle\sigma_{\theta}\rangle = 4.1\degr$. 
The corrected angle dispersion value ($\delta\theta'$), to be used in Equation \ref{e:chandfermi},
may be obtained considering that $\delta\theta = \sqrt{(\delta\theta')^{2}+\langle\sigma_{\theta}\rangle^{2}}$, 
providing $\delta\theta' = 2.9\degr$.  

The velocity dispersion values toward dense fragmentation p5 (which is separated from the selected 
area by only $\approx3$ arcmin) was obtained by \citet{urquhart2009} 
through the study of $^{12}$CO and $^{13}$CO lines, providing 
respectively $\Delta V = 2.0$ and $1.9$ km s$^{-1}$. We have used the average value, 
i.e., $\Delta V = 1.95$ km s$^{-1}$. A careful inspection of the extinction levels 
within the polygon indicate a mean value of $A_{V}\approx13$ mag. Using the same 
gas-to-dust relation discussed in Section \ref{s:fragsmall_sh229}, and 
assuming an angular diameter of about $3'$ (corresponding to an interstellar structure 
at the cavity's border with length of $\approx 1$ pc), we find $n_{\mathrm{H_{2}}} \approx 4.0 \times 10^{3}$ cm$^{-3}$.

Applying these quantities to Equation \ref{e:chandfermi} provides $B\approx400 \,\mu$G,
which is a much larger value as compared to the magnetic field strength in the diffuse 
interstellar medium \citep[$\sim 6\,\mu$G, ][]{beck2001,heiles2005}. This evidence
corroborates the previous hypothesis that a higher $B$ level should arise within this 
area, as a consequence of the cavity's expansion, leading to a pilling up effect. 
Magnetic field estimates toward star-forming regions through the CF method, as well as
Zeeman effect analyses, show that high $B$ values are typical in such environments. 
On one hand, values ranging from $80\,\mu$G \citep[for example, toward the pre-stellar core L 183, ][]{crutcher2004} 
up to about $50000 - 80000\,\mu$G have already been found 
\citep[see the study of W 3, Orion KL, W49N, and S140 regions, conducted by ][]{fiebig1989}.
On the other hand, much lower values have been found toward the magnetically dominated 
Pipe Nebula \citep[varying between $17$ and $65\,\mu$G, depending on the position 
along the cloud, as discussed by][]{alves2008,franco2010}, which shows several 
starless and quiescent cores. Perhaps this could indicate magnetic field differences 
between high and low mass star-forming regions, although more studies are necessary.

It is important to point out that this $B$ computation is highly approximate, 
due to its dependence on several factors which may only be roughly estimated. However, 
it is still a valid approach that allows to evaluate its order of magnitude.
The low precision is mainly a consequence of the uncertainty related to the interstellar 
density, since this property is dependent on various assumptions, such as: 
the typical value of $R=3.09$ related to the extinction law; the presence of other 
species besides molecular hydrogen, such as atomic and ionized hydrogen; 
the estimate of the cloud line-of-sight thickness, etc. 

\subsection{The distribution of the ratio of total-to-selective extinction values toward Sh 2-29}
\label{s:variationsR_sh229}

In Section \ref{s:polserk_sh229} we have shown how the multi-band polarimetric analysis 
could be used to derive the $R$ values to 87 objects from our sample. As discussed before, 
this parameter is highly influenced by the interstellar grain size distribution, and 
therefore may be used as a probe of the dust conditions surrounding the central cavity.
The typical grain size at the diffuse interstellar medium corresponds to 
$\langle a \rangle \approx 0.15 \,\mu$m, which represents $R\approx 3.09$. 
Considering a diffuse cloud with a standard grain size distribution composed by 
particles larger and smaller than the mean value, the $R$ value is proportional 
to the ratio between the number of larger and smaller grains. Therefore, an increase
in the relative number of larger grains will generate a higher $R$ value, and vice-versa.

   \begin{figure}[!t]
   \centering
   \includegraphics[width=0.48\textwidth]{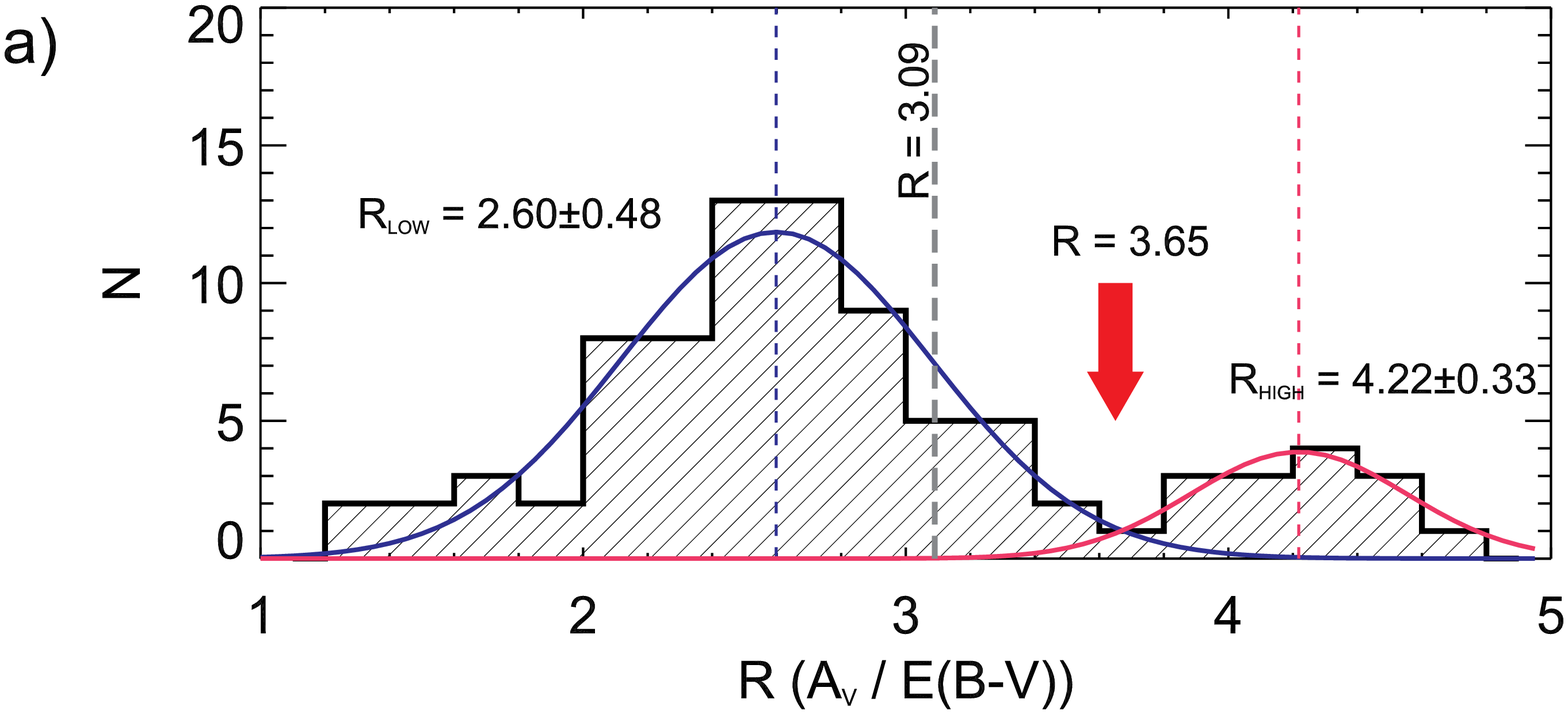} \\
   \includegraphics[width=0.48\textwidth]{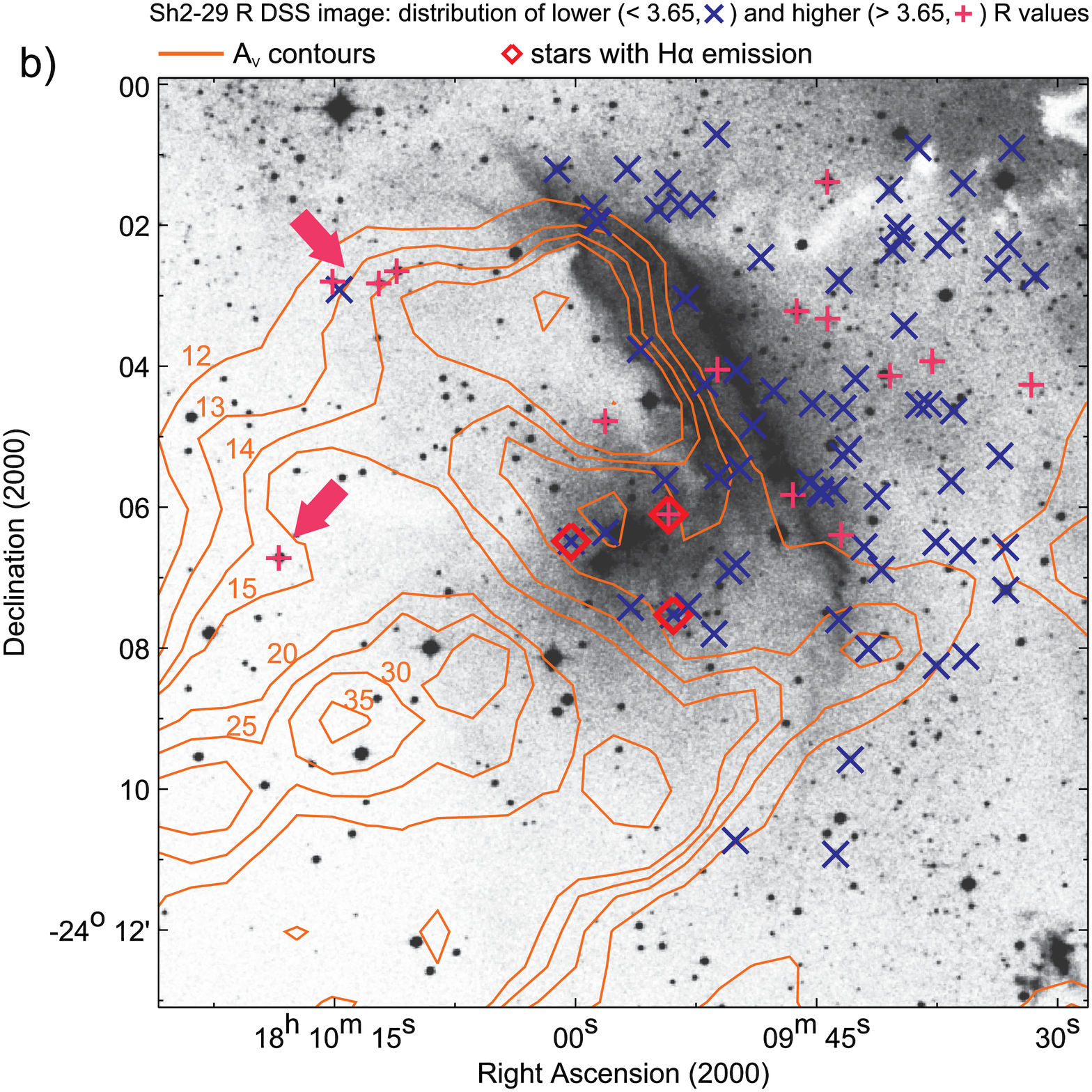} \\
      \caption{Diagram ({\it a}) shows the distribution of $R$ values, using the 
               results of the 87 Serkowski fits computed from the multi-band polarimetric sample. Solid 
               curves represent Gaussian fits to the higher (red line) and lower (blue line) 
               $R$ distribution trends,
               respectively centered at $2.60\pm0.48$ and $4.22\pm0.33$ 
               (the uncertainties correspond to the 1-$\sigma$ dispersion). 
               Diagram ({\it b}) shows the spatial distribution of objects with  
               $R<3.65$ ($\times$ blue symbol) and $R>3.65$ ($+$ pink symbol). 
               Red diamonds indicate the position of 3 stars with H$\alpha$ emission \citep{ogura2002}. 
              }
         \label{Rhistspatial_sh229}
   \end{figure}

In order to study the distribution of $R$ values toward the Sh 2-29 central area, 
we have built the histogram shown in Figure \ref{Rhistspatial_sh229}a. A concentration peaked 
around the standard value ($R=3.09$) would be expected for the diffuse interstellar 
medium \citep[][Fig. 7]{larson2005}.
Instead, it is clear from this diagram that two opposite trends dominate the distribution, 
both diverging considerably from the standard value. The most prominent concentration 
corresponds to a pronounced peak around $R_{low}=2.60\pm0.48$ (as indicated by the blue
Gaussian fit), i.e., showing a predominance of $R$ values lower than $3.09$. 
The second trend is related to higher values, peaked around $R_{high}=4.22\pm0.33$ 
(as shown by the red Gaussian fit). It is less marked as compared to the first trend, 
suggesting that $R$ values toward the central cavity are predominantly lower
than $3.09$, although some few lines-of-sight indicate higher values.

In order to study the different properties between both trending populations of $R$ values, 
we notice from Figure \ref{Rhistspatial_sh229}a that these distributions are well separated, 
in a sense that there is not much superposition between them. In fact, we may 
define the value $R=3.65$ (indicated by the red arrow) as a convenient limiting value 
roughly distinguishing both distributions.

Figure \ref{Rhistspatial_sh229}b shows the spatial distribution of $R$ values larger 
($+$ symbol, pink color) and smaller than $3.65$ ($\times$ symbol, blue color), toward the 
central cavity. There is obviously a mixture of directions with $R>3.65$ and $R<3.65$, 
with a clear predominance of smaller values, specially toward the ionization front. 
It is notable, however, that examining the regions to the left of the central cavity 
(where the denser extension of the dark cloud is located), 4 among 5 of the objects 
with $R$ values present $R>3.65$ (as indicated by the upper and lower pink arrows).
To the only star with $R<3.65$ in this area (located immediately below the upper arrow), 
the computed value is $R=3.2$, which is still above the standard $3.09$ value.
Therefore, there is a weak evidence that regions with higher $R$ values suffered lower
impact from the massive central stars, being located near the edges of the cloud's denser portions, 
therefore probably shielded and protected from the action of intense shock fronts.
Obviously, several stars with $R>3.65$ may also be observed toward the central cavity 
and surrounding its prominent arc-shaped ionization front. However, it is important 
to point out that its individual projected distances are not known, and therefore, these objects 
may also be located at the cloud's nearer edge tracing the neutral instead of ionized gas.

   \begin{figure}[!t]
   \centering
   \includegraphics[width=0.48\textwidth]{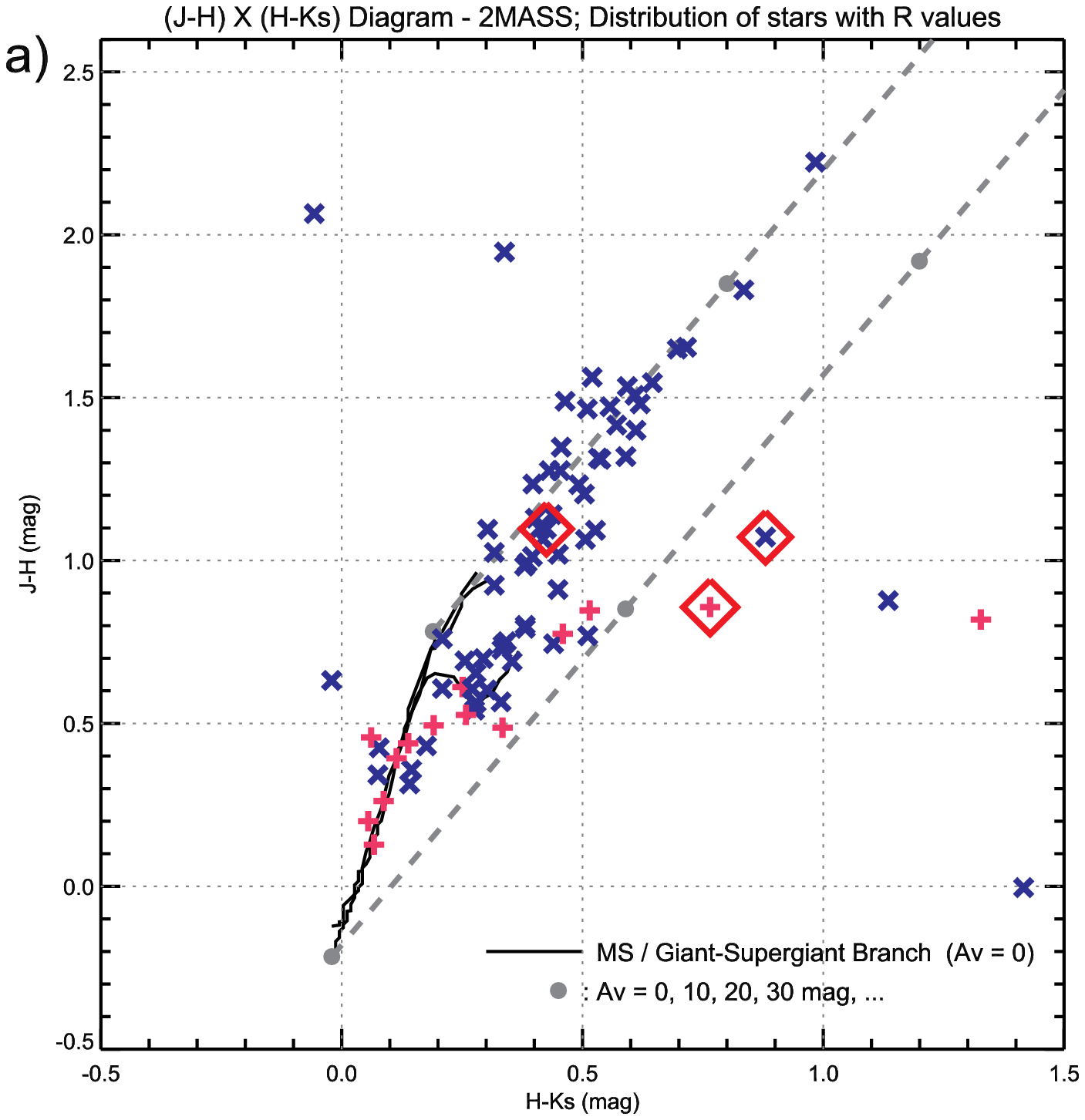} \\
   \includegraphics[width=0.48\textwidth]{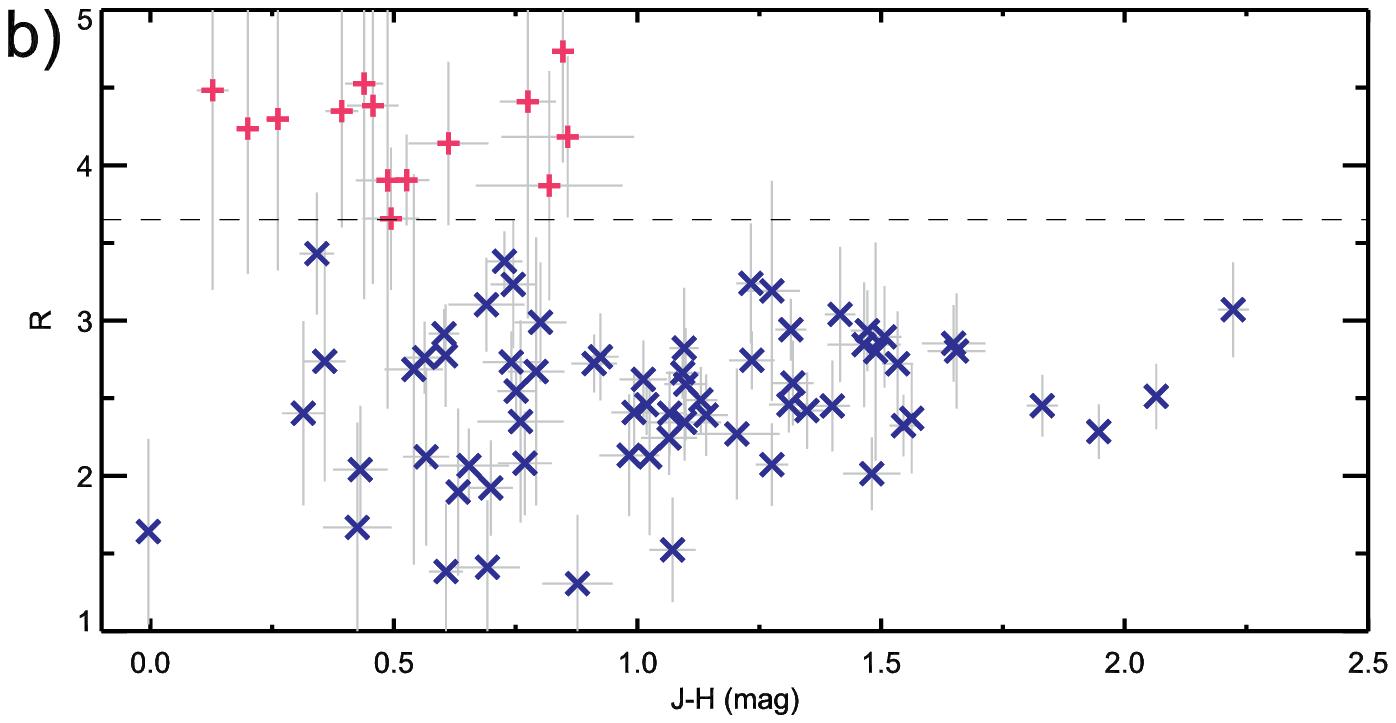} \\
      \caption{({\it a}) Color-color diagram ($J-H \times H-Ks$) for objects with $R$ values, based 
               on 2MASS magnitudes. Symbols and colors follow the same criteria used in Figure 
               \ref{Rhistspatial_sh229}b. The positions of the 
               un-reddened main sequence, giants and super-giants, as well as the reddening band, 
               are similar to Figure \ref{forepol_sh229}a. 
               ({\it b}) $R \times J-H$ diagram, using the same stars from part ({\it a}). 
               The horizontal dashed line corresponds to $R=3.65$, as defined in order to 
               separate between the higher and lower $R$ values. 
              }
         \label{Rcc_sh229}
   \end{figure}

A further analysis may be carried out, by correlating the $R$ values with 2MASS 
near-IR photometric data. Figure \ref{Rcc_sh229}a shows the color-color diagram
($J-H \times H-Ks$) with different symbols and colors for stars with 
$R<3.65$ and $R>3.65$, as in Figure \ref{Rhistspatial_sh229}b. Objects 
with $R<3.65$ are predominantly distributed along the reddening band, 
presenting a wide variety of extinction levels (between $0$ and $\approx15$ mag). 
On the other hand, most of the objects with $R>3.65$ are concentrated 
near the un-reddened main sequence locus, with an exception of only two objects
showing color excess in the near-IR (located to the right of the reddening band).
Such trends are even more evident in the $R \times J-H$ diagram shown in 
Figure \ref{Rcc_sh229}b. Notice that for objects with $R>3.65$, there is 
a limiting $J-H$ value of roughly $0.9$ mag, suggesting a lower reddening 
levels toward these stars. This is a further evidence that objects with higher $R$ values,
even those which are located in the direction of the arc-shaped ionization front,
are possibly located at the edges of the cloud encompassing the star-forming region, 
where $A_{V}$ levels are lower.

\subsection{Interpreting the anomalous $R$ values toward Sh 2-29: processing of interstellar dust grains}

Several processes in the interstellar medium may cause alterations in the grain sizes, 
leading to anomalous $R$ values, which may be higher \citep{vrba1985,whittet2001}
or lower \citep{larson1996,szomoru1999,larson2005} than the standard value, 
depending on the environment's local physical conditions \citep{jones2004,mazzei2008}. 


Normally, larger $R$ values may be found within dense clouds, where coagulation effects 
due to grain-grain collisions are common. However, coagulation will only be sufficiently efficient 
on low-velocity collisions (between $0.001$ and $0.02$ km s$^{-1}$), which 
implies on relatively isolated clouds, generally unaffected or shielded from turbulence 
or shock waves \citep{chokshi1993,jones1996,hirashita2012,kohler2012}. 
Other process that could lead to higher $R$ values is the grain evaporation (or sublimation) 
due to the incidence of highly energetic photons (extreme ultraviolet and $\gamma$-rays), 
since smaller grains are most sensitive to this kind of grain destruction mechanism 
\citep{guhathakurta1989}.

On the other hand, smaller $R$ values are usually related to regions affected by 
shock waves, where sputtering effects (due to the collision between grains and atoms/ions)
tend to favor the destruction of larger grains \citep{mathis1977,jones1996}.
Other possibility is that the turbulence generated by shock waves lead to a higher 
mean grain velocity, which results in fragmentation effects: it is know that for 
velocities higher than $\approx 1$ km s$^{-1}$, coagulation effects will be replaced by 
fragmentation of dust particles, which constitutes an efficient mechanism to convert 
grains from large into smaller sizes \citep{borkowski1995,jones1996}.

\citet{mccall1990} found an anomalous value of $R=4.6\pm0.3$ toward the Lagoon region, 
which is spatially close to Sh 2-29. According to their interpretation, 
stars from the sample used in that case are most likely related to neutral gas, 
and therefore are probably located at the H{\sc ii} region's edges.
It was proposed that the higher $R$ value is probably due to a combined 
effect of coagulation and evaporation by energetic photons which traverses the cloud's dense and 
neutral external shells. In the case of the coagulation hypothesis
the anomaly would be the result of grain growth effects due to the presence of a relic component
of the original dense cloud which gave birth to the star-forming region.

In view of the several physical effects which may alter the average grain sizes, 
the existence of two opposite grain populations toward Sh 2-29 may now be interpreted. 
In Section \ref{s:variationsR_sh229} we have shown that the higher $R$ component 
seems to be related to peripheral locations (near the outskirts of the dense cloud), 
as well as to lower extinction values. Therefore, these sites are possibly remnants 
of the initial dense cloud which originated the star-forming region, probably 
still weakly disturbed by star formation effects, such as shock waves due to massive 
stars. At such relic cloud component, coagulation effects probably dominated 
the previous dense environment, leading to larger $R$ values. Therefore, this population
of large grains is probably still preserved in this environment which 
is seemingly protected from the grain destruction mechanisms. 
This effect is very similar to the one detected toward the Lagoon
nearby region, corroborating our hypothesis. 
Some stars with higher $R$ values could actually be regarded as foreground objects, although it would 
not be consistent with the fact the foreground component is probably represented by a standard
extinction law, since it corresponds to the diffuse interstellar medium.

On the other hand, the lower $R$ component is predominant inside the central cavity
and toward the arc-shaped H$\alpha$ feature, where there is certainly a large influence
of expanding ionization fronts, turbulence and stellar winds from young massive stars.
This environment is consistent with sputtering and grain fragmentation effects, which 
promotes an efficient mechanism that converts larger into smaller dust particles, 
leading to lower mean $R$ values. 

Concluding, the observation of different grain populations toward a single region 
from the interstellar medium is remarkable, revealing the evolution process of 
dust particles. In this case the evidence suggest that the grains' evolutionary stage
surrounding Sh 2-29 corresponds to a gradual conversion 
from an earlier predominantly larger size component to an environment where smaller size dust 
particles now prevail. Such processing effects of interstellar dust, observed during the  
course of the particles' transformation mechanism, has rarely been detected before within
a specific region of the Galaxy, and lends support to the theories of grain evolution.

\section{Conclusions}
\label{s:conclusionssh229}

In this work we have used multi-band polarimetric data (V, R, I and H) toward Sh 2-29
in order to study properties from the interstellar magnetic field lines, as well 
as the size distribution of dust particles. A visual extinction map, 
together with R-band (DSS) and mid-IR ({\it Spitzer}) images were used to reveal 
several interstellar features. The main results from this analysis are listed 
below. 

\begin{enumerate}

\item The most striking feature from Sh 2-29 is an interstellar cavity surrounding an
embedded stellar cluster. The analysis of R and H-band polarimetric mappings, together 
with the visual extinction levels, suggests that this cavity is expanding and dragging 
both the interstellar material and the magnetic field lines outwards, which pile-up 
along its arc-shaped borders. This effect probably leads to a higher magnetic field strength
of about $400 \,\mu$G, as estimated from the Chandrasekhar-Fermi method. Consequently, 
grain alignment mechanisms are more efficient, resulting in higher levels of 
polarization degree, as revealed by the study of polarization as a function of 
radial distance from the center;

\item After correcting the R-band polarization sample from the foreground component, 
statistical analyses were performed by computing the angular dispersion function,
in order to evaluate the magnetic turbulence degree ($\langle B_{t}^{2}\rangle^{1/2}/B_{0}$) 
toward several areas distributed throughout the H{\sc ii} region. In a general way, 
lower turbulence degree values ($30-35\%$) are found toward areas that are probably 
undergoing an expansion process due to the ionized gas, hence leading to a highly
ordered pattern characterizing the compressed field lines. On the other hand, 
areas unaffected by the expansion, located outside the H{\sc ii} region toward its South, 
generally present higher turbulence degree levels ($48-51\%$).
An exception occurs toward the eastern portion of the dark cloud, where its arc-shaped extension
seems to have compressed field lines located immediately below its southern rim;
 
\item A group of prominent, dust extinction peaks, were found to exist surrounding the central cavity's external
borders (probably generated due to its expansion), 
with peak visual extinction values between $20$ and $37$ mag. Estimated levels
of molecular hydrogen density and mass are respectively in the ranges of $0.7$ -- $1.7 \times10^{4}$
cm$^{-3}$ and $46$ -- $112$ solar masses, which is compatible with interstellar clumps.
Their positions are correlated with point-like sources from {\it Spitzer}'s $24\,\mu$m observations, 
and therefore likely represent new generations of young stars being born within its 
densest portions, showing sufficient mid-IR emission from a warm circumstellar disk.
Interpretation of the magnetic field configuration, correlated with the structures' density and extinction 
levels, suggest that the cavity's expansion may have been hindered (due to magnetic pressure) 
or facilitated, depending on the relation between the expansion direction and field line orientation;

\item The multi-band polarimetric analysis allowed the study of the total-to-selective 
extinction distribution, revealing two clearly distinct trends peaked at 
$R_{low}=2.60\pm0.48$ and $R_{high}=4.22\pm0.33$, i.e., both diverging significantly from 
the standard $3.09$ value. The first trend is predominant, and is probably related to lower
mean grain sizes inside the central cavity and surroundings due to intense shock fronts, leading
to sputtering and dust fragmentation effects. The second trend is less marked, and is seemingly
related to a relic component of the previous dense cloud which gave birth to the star-forming site, 
where probably coagulation effects resulted in larger mean grain sizes. 
\end{enumerate}

Sh 2-29 has been revealed as a very rich Galactic environment, where several interesting 
aspects of star formation sites may be simultaneously observed: 
young stars and its interaction with the surrounding environment, disturbance of
interstellar magnetic fields, triggered formation of dense clumps, processing 
of interstellar dust particles, etc. Future studies will involve a near-IR photometric and spectroscopic 
analysis of the embedded stellar population, which is pending a detailed characterization. 
Moreover, a complete characterization of the dense fragmentations (with sub-mm and radio wavelengths)
could reveal interesting properties of the next young stellar generation. 

\acknowledgements
We are grateful to the anonymous referee for the very valuable suggestions and comments.

We thank the staff of the CTIO (Chile) and OPD/LNA (Brazil) for their hospitality and 
invaluable help during our observing runs. 

This work was partially supported by the Department of Physics of the 
Universidad de La Serena. ARL thanks the financial support from Diretoria de 
Investigaci\'on -- Universidad de La Serena through project DIULS REGULAR PR13144.

CRZ acknowledges support from project CONACYT 152160, Mexico.

This investigation made extensive use of data products from the Two Micron All Sky Survey (2MASS), 
which is a joint project of the University of Massachusetts and the Infrared Processing and 
Analysis Center/California Institute of Technology, funded by the National Aeronautics and 
Space Administration and the National Science Foundation.  

This research is based in part on observations made with the Spitzer Space Telescope, which 
is operated by the Jet Propulsion Laboratory, California Institute of Technology under a 
contract with NASA.

We are grateful to Drs. A. M. Magalh\~aes and A. Pereyra for providing the polarimetric unit 
and the software used for data reductions. This research has been partially supported by the
Brazilian agencies FAPEMIG, CNPq and CAPES.

{\it Facilities:} 
\facility{CTIO: 0.9\,m, LNA: 1.6\,m, LNA: 0.6\,m}

\bibliography{astroref}
\bibliographystyle{aa}

\end{document}